\documentclass[10pt, journal,twoside]{IEEEtran}
\usepackage{subfigure}
\usepackage{amsmath}
\usepackage{amssymb}
\usepackage{amsfonts}
\usepackage{amscd}
\usepackage{mathrsfs}
\usepackage{graphicx}
\usepackage{graphics}
\usepackage{color}
\usepackage{url}
\usepackage{bm}
\usepackage{algorithm}
\usepackage{algorithmic}
\usepackage{setspace}

\newtheorem{theorem}{Theorem}
\newtheorem{lemma}{Lemma}
\newtheorem{definition}{Definition}
\newtheorem{proposition}{Proposition}

\newtheorem{remark}{Remark}

\newcounter{MYtempeqncnt}

\newcommand{\samp}{\hookleftarrow}

\newcommand{\mR}{\mathbb{R}}

\newcommand{\poly}{\mathrm{poly}}
\newcommand{\eps}{\varepsilon}

\newcommand{\aln}[1]{\ensuremath{\mathsf{#1}}}
\begin{document}
\title{Comments on ``Physical-layer cryptography through massive \aln{MIMO}''}
\author{Amin Sakzad and Ron Steinfeld$^\dagger$
\thanks{$^\dagger$ Amin Sakzad and Ron Steinfeld are with
Faculty of Information Technology,
Department of Software Systems and Cybersecurity, 
Monash University, Clayton VIC 3800, Australia.
E-mail: $\tt \{amin.sakzad,ron.steinfeld$\}@$\tt monash.edu$. Amin Sakzad and Ron Steinfeld were both supported by the Australian Research Council (ARC) under Discovery grants ARC DP~$150100285$. A subset of this work was presented in~\cite{SS15c} at ITW $2015$, Jeju Island, South Korea.}
}

\maketitle
%\IEEEpeerreviewmaketitle
%%%%%%%%%%%%%%%%%%%%%%%%%%%%%%%%%%%%%%%%%%%%%%%%%%%%%%%%%%%%%%%%%%%%%%%%%%%%%%%%%%%%%%%%%%%%%%%%%%%%%%%%%%%%%%%%
%%%%%%%%%%%%%%%%%%%%%%%%%%%%%%%%%%%%%%%%%%%%%%%%%%%%%%%%%%%%%%%%%%%%%%%%%%%%%%%%%%%%%%%%%%%%%%%%%%%%%%%%%%%%%%%%
\begin{abstract}
We present two attacks on two different versions of physical layer cryptography schemes based on massive multiple-input multiple-output (\aln{MIMO}). Both cryptosystems employ a singular value decomposition (\aln{SVD}) precoding technique. For the first one, we show that the eavesdropper (who knows its own channel and the channel between legitimate users) can decrypt the information data under the same condition as the legitimate receiver. We study the signal-to-noise advantage ratio for decoding by the legitimate user over the eavesdropper in a more generalized scheme when an arbitrary precoder at the transmitter is employed. On the negative side, we show that if the eavesdropper uses a number of receive antennas much larger than the number of legitimate user antennas, then there is no advantage, independent of the precoding scheme employed at the transmitter. On the positive side, for the case where the adversary is limited to have the same number of antennas as legitimate users, we give an $\mathcal{O}\left(n^2\right)$ upper bound on the advantage and show that this bound can be approached using an inverse precoder. For the second cryptosystem, we show that the required security conditions prevent the legitimate user from decoding the plain-text uniquely.
%To complete our study regarding inverse precoding approach, we find the distribution of the singular values of the quotient of two Gaussian matrices. We further investigate the distribution of an orthogonal transformation of such a quotient thorough %\aln{QR} decomposition. This gives us the opportunity to lower bound the decoding advantage ratio of the legitimate user over an eavesdropper who is equipped with a non-linear successive interference cancellation (\aln{SIC}) stronger than linear receivers.
\end{abstract}
%%%%%%%%%%%%%%%%%%%%%%%%%%%%%%%%%%%%%%%%%%%%%%%%%%%%%%%%%%%%%%%%%%%%%%%%%%%%%%%%%%%%%%%%%%%%%%%%%%%%%%%%%%%%%%%%
%%%%%%%%%%%%%%%%%%%%%%%%%%%%%%%%%%%%%%%%%%%%%%%%%%%%%%%%%%%%%%%%%%%%%%%%%%%%%%%%%%%%%%%%%%%%%%%%%%%%%%%%%%%%%%%%
\begin{IEEEkeywords}
Physical Layer Cryptography, Massive \aln{MIMO}, Precoding, Zero-Forcing Linear Receiver.%, Successive Interference Cancelation.
\end{IEEEkeywords}
%%%%%%%%%%%%%%%%%%%%%%%%%%%%%%%%%%%%%%%%%%%%%%%%%%%%%%%%%%%%%%%%%%%%%%%%%%%%%%%%%%%%%%%%%%%%%%%%%%%%%%%%%%%%%%%%
%%%%%%%%%%%%%%%%%%%%%%%%%%%%%%%%%%%%%%%%%%%%%%%%%%%%%%%%%%%%%%%%%%%%%%%%%%%%%%%%%%%%%%%%%%%%%%%%%%%%%%%%%%%%%%%%
\section{Introduction}
%%%%%%%%%%%%%%%%%%%%%%%%%%%%%%%%%%%%%%%%%%%%%%%%%%%%%%%%%%%%%%%%%%%%%%%%%%%%%%%%%%%%%%%%%%%%%%%%%%%%%%%%%%%%%%%%
%%%%%%%%%%%%%%%%%%%%%%%%%%%%%%%%%%%%%%%%%%%%%%%%%%%%%%%%%%%%%%%%%%%%%%%%%%%%%%%%%%%%%%%%%%%%%%%%%%%%%%%%%%%%%%%%
\emph{Background.} Since the pioneering theoretical study of the ``wiretap channel'' by Wyner~\cite{Wy75}, various techniques for achieving secure communication have been proposed based on physical assumptions on the communication channel. These methods, known as ``physical layer security'', ensure that the communication channel between the legitimate parties is sufficiently ``different'' from the channel between the legitimate parties and the adversaries. Since such methods do not assume an existing shared secret key between legitimate parties, nor require the secure storage of any secret key, they offer a potential physical alternative in some applications to classical software-based cryptographic techniques such as public-key cryptography~\cite{DH76}. In the context of wireless communications, such methods have the novel feature of replacing the role of the secret key needed for decryption in classical cryptosystems, with the \emph{physical location} of the legitimate receiver's antennas, so that security should be achieved against an adversary whose antennas are located in a sufficiently different location (the difference in location typically need only be significant with respect to the signal wavelength; thus for microwave communication, only a very small distance would already guarantee security). Unfortunately, to achieve their information-theoretic security properties, most existing physical layer security techniques need to assume significant additional limitations on the resources or capability of the adversary, which may not be realistic in many practical applications; for example, the techniques in~\cite{OH11} assume that the signal-to-noise ratio in the adversary's channel is smaller than the signal-to-noise ratio in the legitimate receiver channel, while \aln{MIMO} ``jamming'' techniques such as those based on ``artificial noise''~\cite{GN08,ZSB16} need to assume that the number of receiving antennas used by the adversary $n'_r$ is smaller than the number of transmitting antennas $n_t$ or the number of receiving antennas $n_r$ of the legitimate sender and receiver, respectively.

Recently, an interesting new approach for physical security in massive multiple-input multiple-output (\aln{MIMO}) communication systems was introduced by Dean and Goldsmith~\cite{DG13v1,DG13v2} and called ``Physical layer cryptography'', or a massive \aln{MIMO} physical layer cryptosystem (\aln{MMPLC}). In this scenario, the channel state information (\aln{CSI}) is known at the legitimate transmitter as well as all the other adversaries and legitimate receivers. The eavesdropper has also the knowledge of the \aln{CSI} between legitimate users. To achieve such a goal, the authors of~\cite{DG13v1,DG13v2} precode the information data at the transmitter, based on the known \aln{CSI} between the legitimate users, so that the decoding of the received vector would be computationally easy for the legitimate user but computationally hard for the adversary. The above assumptions on the channel conditions seem to be deliberately created for \aln{MMPLC} and not raised naturally from the physical of the channels. First, use of \aln{SVD} beamforming with a constellation with the same spacing between the constellation points does not appear to be technically sound, given that perfect \aln{CSI} is available at the transmitter. Second, the asymptotic in $n_t$ and/or $n_r$ with perfect \aln{CSI} is of no interest, because even if a system with ever increasing number of antennas could be built, finite channel coherence will limit the number of dimensions that can be trained and eventually break the perfect-\aln{CSI} assumption.

The main idea in~\cite{DG13v1,DG13v2} is to replace the information-theoretic security guarantees of previous physical layer security methods with the weaker complexity-based security guarantees used in cryptography. More precisely, the goal of~\cite{DG13v1,DG13v2} is to show that the adversary cannot decode the sent message (using efficient ``Signal Processing'' techniques) due to computational complexity barriers associated to the available massive \aln{MIMO} decoding algorithms. This approach trades-off a weaker, but still practical, complexity-based security guarantee in order to avoid the less practical additional assumptions required by existing information-theoretic techniques, such as stronger noise level in~\cite{OH11,ZSB13,ZSB14,WLWQ14,WLWQ15} and/or less antennas for the adversary than for legitimate parties in~\cite{GN08,ZSB16}, while still retaining the ``no secret key'' location-based decryption feature of physical-layer security methods. For a survey on physical layer security for massive \aln{MIMO} see~\cite{KZR15}.

In~\cite{DG13v1}, a \aln{MMPLC-13} is presented that is claimed to achieve the above goal of the complexity-based approach, using a singular value decomposition (\aln{SVD}) precoding technique and $m$-PAM constellations at the transmitter. Namely, it is claimed that, under a certain condition on the number $n_t$ of legitimate sender's transmit antennas and the noise level $\beta$ in the adversary's channel (which we call the \emph{hardness condition} of \aln{MMPLC-13}), the message decoding and distinguishing problems for the adversary (eavesdropper), termed the \aln{MIMO-Search} and \aln{MIMO-Decision} problem in~\cite{DG13v1}, respectively, are as hard to solve on average as it is to solve a standard conjectured hard lattice problem in dimension $n_t$ in the worst-case, in particular, the $\aln{GapSVP}_{\poly(n_t)}$ variant of the approximate shortest vector problem in arbitrary lattices of dimension $n_t$, with approximation factor polynomial in $n_t$. For these problems, no polynomial-time algorithm is known, and the best known algorithms run in time exponential in the number of transmit antennas $n_t$, which is typically infeasible when $n_t$ is in the range of few hundreds (as in the case of massive \aln{MIMO}). Significantly, this computational hardness of \aln{MIMO-Decision} is claimed to hold even if the adversary is allowed to use a large number of receive antennas $n'_r = \poly(n_t)$ \emph{polynomially larger} than $n_t$ and $n_r$ used by the legitimate parties, and with the same noise level as the legitimate receiver ($\beta=\alpha$). Consequently, under the widely believed conjecture that no polynomial-time algorithms for $\aln{GapSVP}_{\poly(n_t)}$ in dimension $n_t$ exist and the hardness condition of~\cite{DG13v1}, the authors of~\cite{DG13v1} conclude that their \aln{MMPLC-13} and the corresponding \aln{MIMO-Decision} problem is secure against adversaries with run-time polynomial in $n_t$.

In~\cite{DG13v2}, \aln{MMPLC-17} is provided, which is basically same as \aln{MMPLC-13} and claimed to achieve the complexity-based security based on a weaker hardness assumption and different security conditions. In particular, it is shown that, under two certain conditions (different from that in \aln{MMPLC-17}) on the number $n_t$ of legitimate sender's transmit antennas, the number $n_r$ of legitimate user's receive antennas, and the constellation size $m$, the message decoding problem (the \aln{MIMO-Search} problem in~\cite{DG13v2}) for the adversary, is as hard to solve on average as (above mentioned) lattice problems in dimension $n_t$ in the worst-case. We call the latter two conditions, the \emph{hardness conditions of \aln{MMPLC-17}}. Note that there are two differences between \aln{MMPLC-13} and \aln{MMPLC-17}: ({\em i}) first there is only one hardness condition in \aln{MMPLC-13}, while there are two other hardness conditions in \aln{MMPLC-17} both different from \aln{MMPLC-13}, ({\em ii}) the cryptosystem in~\cite{DG13v1} is claimed to be secure since both \aln{MIMO-Search} and hence \aln{MIMO-Decision} are hard, but the security of the scheme in~\cite{DG13v1} is base on the hardness of \aln{MIMO-Search} only.

\emph{Our Contribution.} In this paper, we further analyse the complexity-based \aln{MMPLC-13} and \aln{MMPLC-17} initiated in~\cite{DG13v1,DG13v2}, to improve the understanding of their potential and limitations. Our contributions are summarized below:
\begin{itemize}
\item{\bf Security of \aln{MMPLC-13} is flawed.} Using a linear receiver known as zero-forcing (\aln{ZF})~\cite{Kumar09}, a well-known and efficient Signal-Processing algorithm with run-time polynomial in $n_t$, we show that \aln{MIMO-Search} problem defined in~\cite{DG13v1} can be solved efficiently under an extra condition on the number of receive antennas. We analyse the decoding success probability of this algorithm and prove that it is $\geq 1-o(1)$ even if the \emph{hardness condition} of \aln{MMPLC-13} is satisfied, if the ratio $y'=n_r'/n_t$ exceeds a small factor at most logarithmic in $n_t$, i.e. $y' = \mathcal{O}(\log n_t)$ asymptotically. This contradicts the hardness of the \aln{MIMO-Search} problem conjectured in~\cite{DG13v1} to hold for much larger polynomial ratios $y' =\mathcal{O}(\poly(n_t))$. Note the number of transmit antennas $n_t$ is considered as the security parameter of \aln{MMPLC-13}, and hence the number of receive antennas in the employed massive \aln{MIMO} is in the order of few hundreds. This justifies the reason why we derived and discussed asymptotic results on \aln{MMPLC-13}. Moreover, we show that the decoding success probability of an adversary against the \aln{MMPLC-13} of~\cite{DG13v1} using the \aln{ZF} decoder is approximately the same (or greater than) as the decoding success probability of the legitimate receiver using a maximum-likelihood \aln{ML} decoder if $n'_r$ is approximately greater than or equal to $n_r$, assuming an equal noise level for adversary and legitimate receivers. Our first contribution implies that the \aln{SVD} precoder-based \aln{MM-PLC} in~\cite{DG13v1} still requires for security an undesirable assumption limiting $n'_r$ to be less than that of the legitimate receiver, similar to previous information-theoretic techniques.

\item{\bf \aln{MMPLC-17} is cryptographically incorrect.} We show that, by combining the two hardness conditions of \aln{MMPLC-17} in~\cite{DG13v2} for $n_t$, $n_r$, and $m$, we derive a new condition (based upon only $n_t$ and $m$) which implies that the legitimate user cannot uniquely decode the sent message independent of its updated security argument compared to \aln{MMPLC-13}. In particular, if ${\bf x}$ is sent, we show that the legitimate user can not uniquely decode to ${\bf x}$, as ${\bf x}+{\bf e}_1$, where ${\bf e}_1$ denote the unit vector with a single $1$ in the first coordinate and $0$ elsewhere, is statistically close to ${\bf x}$.

\item{\bf Potential of \aln{MMPLC}.} As last contribution, we investigate the potential of the general approach of~\cite{DG13v1} and~\cite{DG13v2} by studying the generalized scenario where one allows arbitrary precoding matrices by the legitimate transmitter in place of the \aln{SVD} precoder. To do so, we define a decoding advantage ratio for the legitimate user over the adversary, which is approximately the ratio of the maximum noise power tolerated by the legitimate user's decoder to the maximum noise power tolerated by the adversary's decoder (for the same ``high'' success probability). We derive a general upper bound on this advantage ratio, and show that, even in the general scenario, the advantage ratio tends to $1$ (implying no advantage), if the ratio $n'_r/\max(n_t,n_r)$ exceeds a small constant factor ($\leq 9$). We further show that user $\mathrm{B}$ has essentially no decoding advantage over user $\mathrm{E}$ when user $\mathrm{E}$ has the same (or bigger) number of receiving antennas. Thus a linear limitation (in the number of legitimate user antennas) on the number of adversary antennas seems inherent to the security of this approach. On the positive side, we show that, in the case when legitimate parties and the adversary all have the same number of antennas ($n'_r=n_r=n_t$), the upper bound on the advantage ratio is quadratic in $n_t$. We give both theoretical and experimental evidences that this upper bound can be achieved using an inverse precoder instead of \aln{SVD} precoder. Notice that, we neither introduce a new precoder (in the sense of Telecommunication theory) nor a new cryptosystem through inverse precoder. Instead, we use this power-inefficient precoder to only show the sharpness/achievability of our bounds on advantage ratio. In particular, we study the distribution of the quotient of two Gaussian matrices and its least singular value. We further derive the distribution of the diagonal elements of an upper triangular matrix obtained in the \aln{QR} decomposition of the mentioned quotient matrix. These results enable us to define and derive explicitly the decoding advantage ratio for the legitimate user over the adversary equipped with a successive interference cancellation \aln{SIC} decoder.
\end{itemize}
\begin{remark}
Note that the first bullet of the above mentioned contributions is also published in~\cite{SS15c}. The second and third contributions in Sections~\ref{SubSection:Discussion} and~\ref{AnUpperBoundonAdvantageRatio}, are completely new compared to what is presented in~\cite{SS15c}.
\end{remark}
{\bf Notation.} The notation $a\gg b$ denotes that the real number $a$ is much greater than $b$. We let $|z|$ denotes the absolute value of $z$. Vectors will be column-wise and denoted by bold small letters. Let ${\bf v}$ be a vector, then its $j$-th entry is represented by $v_j$. A $k_1\times k_2$ matrix ${\bf X}=[{\bf x}_1,\ldots,{\bf x}_{k_2}]$ is formed by joining the $k_1$-dimensional column vectors ${\bf x}_1,\ldots, {\bf x}_{k_2}$. The superscript $^t$ denotes transposition operation. We make use of the standard Landau notations to classify the growth of functions. We say that a function $F(n)$ is $\poly(n)$ if it is bounded by a polynomial in $n$. The notation $\omega(F(n))$ refers to the set of functions (or an arbitrary function in that set) growing faster than $cF(n)$ for any constant $c > 0$. A function $G(n)$ is said negligible if it is proportional to $n^{-\omega(1)}$. If $x$ is a random variable and $E$ is a set, $\mathbb{P}[E]$ denotes the probability of the event ``$x \in E$''. The expected value and variance of a random variable $x$ is denoted by $\mathbb{E}[x]$ and $\mathbb{V}[x]$, respectively. The standard Gaussian distribution on $\mR$ with zero mean and variance $\sigma^2$ is denoted by $\mathcal{N}_{\sigma^2}$. We denote by $w \samp \mathcal{D}$ the assignment to random variable $w$ a sample from the probability distribution $\mathcal{D}$. The statistical distance (\aln{SD}) between distributions $\mathcal{D}_1$ and $\mathcal{D}_2$ over a domain $\mathcal{E}$ is
$$\Delta(\mathcal{D}_1, \mathcal{D}_2) = \frac{1}{2}\int_{\mathcal{E}} |\mathcal{D}_1(x) - \mathcal{D}_2(x)|dx.$$
%%%%%%%%%%%%%%%%%%%%%%%%%%%%%%%%%%%%%%%%%%%%%%%%%%%%%%%%%%%%%%%%%%%%%%%%%%%%%%%%%%%%%%%%%%%%%%%%%%%%%%%%%%%%%%%%
%%%%%%%%%%%%%%%%%%%%%%%%%%%%%%%%%%%%%%%%%%%%%%%%%%%%%%%%%%%%%%%%%%%%%%%%%%%%%%%%%%%%%%%%%%%%%%%%%%%%%%%%%%%%%%%%
\section{System Model}\label{Section:Background}
%%%%%%%%%%%%%%%%%%%%%%%%%%%%%%%%%%%%%%%%%%%%%%%%%%%%%%%%%%%%%%%%%%%%%%%%%%%%%%%%%%%%%%%%%%%%%%%%%%%%%%%%%%%%%%%%
%%%%%%%%%%%%%%%%%%%%%%%%%%%%%%%%%%%%%%%%%%%%%%%%%%%%%%%%%%%%%%%%%%%%%%%%%%%%%%%%%%%%%%%%%%%%%%%%%%%%%%%%%%%%%%%%
We first summarize the notion of real lattices and \aln{SVD} (of a matrix) which are essential for the rest of the paper. A $k$-dimensional {\em lattice} $\Lambda$ with a basis set $\{{\boldsymbol\ell}_1,\ldots,{\boldsymbol\ell}_k\}\subseteq\mathbb{R}^d$ is the set of all integer linear combinations of basis vectors. Let ${\bf L}$ be a matrix with ${\boldsymbol\ell}_m$ as its columns, $1\leq m\leq k$, then ${\bf L}$ is called the {\em generator matrix} of the lattice $\Lambda_{\bf L}$. The determinant of a $\Lambda_{\bf L}$ is defined as
$$\det(\Lambda_{\bf L})\triangleq \sqrt{\det({\bf L}^h{\bf L})},$$
where ${\bf L}^h$ denote the Hermitian transposition of the matrix ${\bf L}$. For any lattice $\Lambda_{\bf L}$, the minimum distance of $\lambda_1(\Lambda_{\bf L})$ is the smallest Euclidean distance between any two lattice points.
Let $s\geq t$, then every matrix ${\bf M}_{s\times t}$ admits a singular value decomposition (\aln{SVD}) ${\bf M}={\bf U}{\bf \Sigma}{\bf V}^t$, where the matrices ${\bf U}_{s\times t}$ and ${\bf V}_{t\times t}$ are two orthogonal matrices and ${\bf \Sigma}_{t\times t}$ is a rectangular diagonal matrix with non-negative diagonal elements $\sigma_1({\bf M})\geq \cdots\geq\sigma_s({\bf M})$. By abusing the notation, we denote the Moore--Penrose pseudo-inverse of ${\bf M}$ by ${\bf M}^{-1}$, that is ${\bf V}{\bf \Sigma}^{-1}{\bf U}^t$, where the pseudo-inverse of ${\bf \Sigma}$ is denoted by ${\bf \Sigma}^{-1}$ and can be obtained by taking the reciprocal of each non-zero entry on the diagonal of ${\bf \Sigma}$ and finally transposing the matrix.

We note that the construction in~\cite{DG13v1} and~\cite{DG13v2} are the same and only the hardness conditions are different. Therefore, we first recall the system model of~\cite{DG13v1,DG13v2} and then present the correctness condition (although not given in either) and finally study the hardness (security) conditions of each separately.
%%%%%%%%%%%%%%%%%%%%%%%%%%%%%%%%%%%%%%%%%%%%%%%%%%%%%%%%%%%%%%%%%%%%%%%%%%%%%%%%%%%%%%%%%%%%%%%%%%%%%%%%%%%%%%%%
%%%%%%%%%%%%%%%%%%%%%%%%%%%%%%%%%%%%%%%%%%%%%%%%%%%%%%%%%%%%%%%%%%%%%%%%%%%%%%%%%%%%%%%%%%%%%%%%%%%%%%%%%%%%%%%%
\subsection{Dean-Goldsmith Model}\label{SubSection:SystemModel}
\begin{figure}[htb]%
  \begin{center}%
\includegraphics[width=5cm]{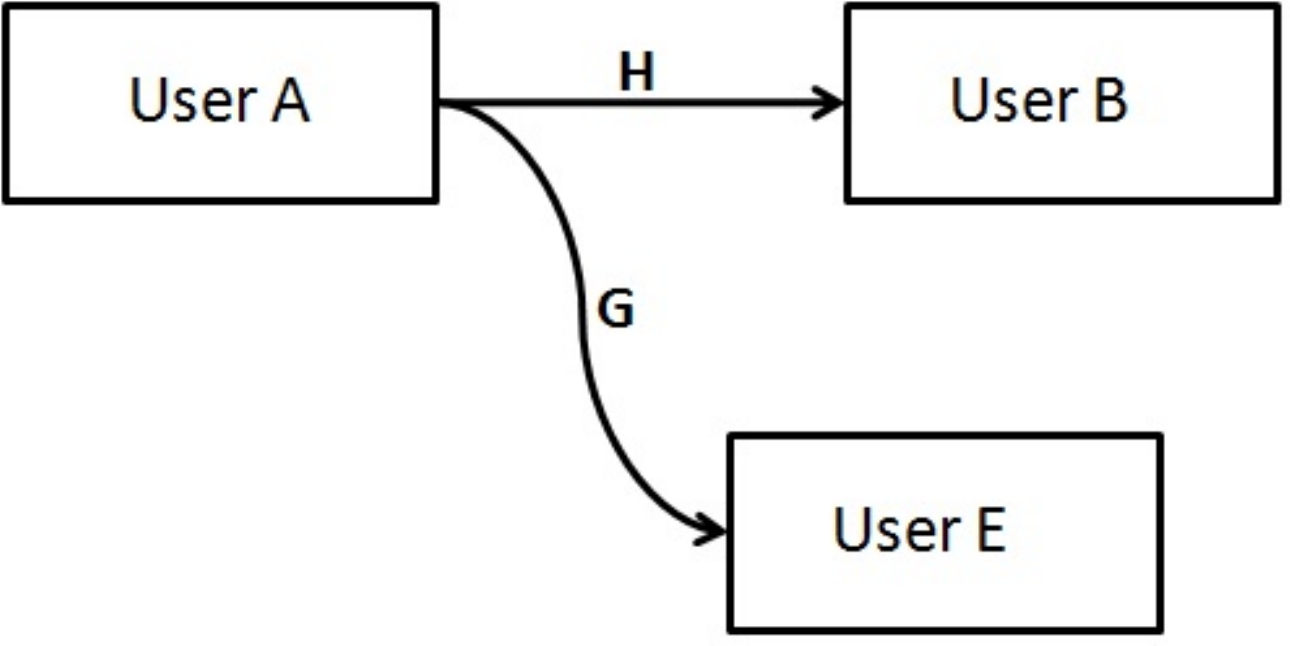}~\caption{\label{fig:MIMOWTC}
The block diagram of a \aln{MIMO} wiretap channel. The channel between user $\mathrm{A}$ and user $\mathrm{B}$ (legitimate users) is denoted by ${\bf H}$. The matrix ${\bf G}$ represents the channel between user $\mathrm{A}$ and the adversary $\mathrm{E}$.}
  \end{center}
\end{figure}

%%%%%%%%%%%%%%%%%%%%%%%%%%%%%%%%%%%%%%%%%%%%%%%%%%%%%%%%%%%%%%%%%%%%%%%%%%%%%%%%%%%%%%%%%%%%%%%%%%%%%%%%%%%%%%%%
%%%%%%%%%%%%%%%%%%%%%%%%%%%%%%%%%%%%%%%%%%%%%%%%%%%%%%%%%%%%%%%%%%%%%%%%%%%%%%%%%%%%%%%%%%%%%%%%%%%%%%%%%%%%%%%%
We consider a slow-fading \aln{MIMO} wiretap channel model as in Fig.~\ref{fig:MIMOWTC}.
The $n_r\times n_t$ real-valued \aln{MIMO} channel from user $\mathrm{A}$ to user $\mathrm{B}$ is denoted by ${\bf H}$. We also denote the channel from $\mathrm{A}$ to the adversary $\mathrm{E}$ by an $n_r'\times n_t$ matrix ${\bf G}$. The entries of ${\bf H}$ and ${\bf G}$ are identically and independently distributed (i.i.d.) based on a Gaussian distribution $\mathcal{N}_1$. We also assume that ${\bf H}$ and ${\bf G}$ are independent as the geographical location of legitimate user and the adversary are different. These channel matrices are assumed to be constant for long time as we employ precoders at the transmitter. This model can be written as:
$$\left\{\begin{array}{l}
{\bf y} = {\bf H}{\bf x} + {\bf e},\\
{\bf y}' = {\bf G}{\bf x} + {\bf e}'.
\end{array}\right.$$
The entries $x_i$ of ${\bf x} \in \mathbb{R}^{n_t}$, for $1\leq i \leq n_t$, are drawn from a constellation $\mathcal{X}=\{0,1,\ldots, m-1\}$ for an integer $m$.
We assume that ${\bf x}$ satisfies an average power constraint $\mathbb{E}(\|{\bf x}\|^2) = \rho$.
The components of the noise vectors ${\bf e}$ and ${\bf e}'$ are i.i.d. based on Gaussian distributions $\mathcal{N}_{m^2\alpha^2}$ and $\mathcal{N}_{m^2\beta^2}$, respectively. We assume $\alpha = \beta\in(0,1)$ to evaluate the potential of the Dean-Goldsmith model to provide security based on computational complexity assumptions, without a ``degraded noise'' assumption on the eavesdropper.
In this communication setup, the \aln{CSI} is available at all the transmitter and receivers. In fact, users $\mathrm{A}$ and $\mathrm{B}$ know the channel matrix ${\bf H}$ (via some channel identification process), while adversary $\mathrm{E}$ has the knowledge of both channel matrices ${\bf G}$ and ${\bf H}$. The knowledge of ${\bf H}$ allows $\mathrm{A}$ to perform a linear precoding to the message before transmission. More specifically, in~\cite{DG13v1,DG13v2}, to send a message ${\bf x}$ to $\mathrm{B}$, user $\mathrm{A}$ performs an \aln{SVD} precoding as follows. Let \aln{SVD} of ${\bf H}$ be given as ${\bf H} = {\bf U}{\bf \Sigma}{\bf V}^t$. The user $\mathrm{A}$ transmits ${\bf V}{\bf x}$ instead of ${\bf x}$ and $\mathrm{B}$ applies a filter matrix ${\bf U}^t$ to the received vector ${\bf y}$. With this, the received vectors at $\mathrm{B}$ and $\mathrm{E}$ are as follows:
$$\left\{\begin{array}{l}
\tilde{\bf y} = {\bf \Sigma}{\bf x} + \tilde{\bf e},\\
{\bf y}' = {\bf G}{\bf V}{\bf x} + {\bf e}',
\end{array}\right.$$
where $\tilde{\bf e}={\bf U}^t{\bf e}$. Note that since ${\bf U}^t$ and ${\bf V}$ are both orthogonal matrices, the vector $\tilde{\bf e}$ and the matrix ${\bf G}_v\triangleq{\bf G}{\bf V}$ continue to be i.i.d. Gaussian vector and matrix, with components of zero mean and variances $m^2\alpha^2$ and $1$, respectively.
%%%%%%%%%%%%%%%%%%%%%%%%%%%%%%%%%%%%%%%%%%%%%%%%%%%%%%%%%%%%%%%%%%%%%%%%%%%%%%%%%%%%%%%%%%%%%%%%%%%%%%%%%%%%%%%%
%%%%%%%%%%%%%%%%%%%%%%%%%%%%%%%%%%%%%%%%%%%%%%%%%%%%%%%%%%%%%%%%%%%%%%%%%%%%%%%%%%%%%%%%%%%%%%%%%%%%%%%%%%%%%%%%
\subsection{Correctness Condition}\label{SubSection:CorrectnessCondition}
%%%%%%%%%%%%%%%%%%%%%%%%%%%%%%%%%%%%%%%%%%%%%%%%%%%%%%%%%%%%%%%%%%%%%%%%%%%%%%%%%%%%%%%%%%%%%%%%%%%%%%%%%%%%%%%%
%%%%%%%%%%%%%%%%%%%%%%%%%%%%%%%%%%%%%%%%%%%%%%%%%%%%%%%%%%%%%%%%%%%%%%%%%%%%%%%%%%%%%%%%%%%%%%%%%%%%%%%%%%%%%%%%
Although Dean-Goldsmith do not provide a correctness analysis in either of ~\cite{DG13v1} and~\cite{DG13v2}, we provide one here for completeness. Since ${\bf \Sigma} = \mbox{diag}(\sigma_1({\bf H}),\ldots,\sigma_{n_t}({\bf H}))$ is diagonal, user $\mathrm{B}$ recovers an estimate $\tilde{x}_i$ of the $i$-th coordinate/layer $x_i$ of ${\bf x}$, by performing two operations dividing and rounding as follows:
$$\tilde{x}_i = \left\lceil \tilde{y}_i/\sigma_i({\bf H})\right\rfloor = x_i + \left\lceil \tilde{e}_i/\sigma_i({\bf H})\right\rfloor.$$
Note that $n_r\geq n_t$, unless otherwise $\sigma_{n_t}({\bf H})=0$. It is now easy to see that the decoding process succeeds if $|\tilde{e}_i| < |\sigma_i({\bf H})|/2$ for all $1\leq i\leq n_t$.
Since each $\tilde{e}_i$ is distributed as $\mathcal{N}_{m^2\alpha^2}$, the decoding error probability, $\mathbb{P}\left[\mathrm{B}|{\bf H}\right]$ that $\mathrm{B}$ incorrectly decodes ${\bf x}$ conditioned on a fixed ${\bf H}$, is, by a union bound, upper bounded by $n_t$ times the probability of decoding error at the worst layer:
\begin{eqnarray}
\mathbb{P}\left[\mathrm{B}|{\bf H}\right]\!\!&\leq&\!\! n_t\mathbb{P}_{w \samp \mathcal{N}_{m^2 \alpha^2}}\left[|w| > |\sigma_{n_t}({\bf H})|/2\right]~\label{Pe:AB0}\\
\!\!&=&\!\!n_t\mathbb{P}_{w \samp \mathcal{N}_{1}}\left[|w| > |\sigma_{n_t}({\bf H})|/(2 m \alpha) \right]~\label{Pe:AB}\\
\!\!&\leq&\!\!n_t \exp\left(-|\sigma_{n_t}({\bf H})|^2/\left(8 m^2 \alpha^2\right)\right),~\label{Pe:AB1}
\end{eqnarray}
where we have used the bound $\exp(-x^2/2)$ on the tail of the standard Gaussian distribution. By choosing $\alpha$ such that
$$\alpha^2 \leq |\sigma_{n_t}({\bf H})|^2/\left(8m^2 \log(n_t/\eps)\right),$$
one can ensure that $\mathrm{B}$'s error probability $\mathbb{P}\left[\mathrm{B}|{\bf H}\right]$ is less than any $\eps>0$.
\begin{remark}
The number of transmit antenna's $n_t$ is defined as the security parameter (commonly used by cryptographers, see~\cite{KatzLindell2007}) in both~\cite{DG13v1} and~\cite{DG13v2}. This means that the system's correctness and security depend asymptotically on $n_t$. In particular, the system is called \emph{correct} if user $\mathrm{B}$ can decode ${\bf x}$ correctly with overwhelming probability $\geq1-n_t^{-c}$, for a positive constant $c$. Furthermore, decoding ${\bf x}$ for $\mathrm{E}$ is hard (or the system is called \emph{computationally secure}) if there exists no efficient decoding algorithm for $\mathrm{E}$, whose its run-time is within some polynomial factor of $n_t$. For more details on the exact definitions of computational correctness and security, please refer to~\cite{KatzLindell2007}.
\end{remark}

%%%%%%%%%%%%%%%%%%%%%%%%%%%%%%%%%%%%%%%%%%%%%%%%%%%%%%%%%%%%%%%%%%%%%%%%%%%%%%%%%%%%%%%%%%%%%%%%%%%%%%%%%%%%%%%%
%%%%%%%%%%%%%%%%%%%%%%%%%%%%%%%%%%%%%%%%%%%%%%%%%%%%%%%%%%%%%%%%%%%%%%%%%%%%%%%%%%%%%%%%%%%%%%%%%%%%%%%%%%%%%%%%
\subsection{Security Condition of the Cryptosystem in~\cite{DG13v1}}\label{SubSection:SecurityCondition1}
%%%%%%%%%%%%%%%%%%%%%%%%%%%%%%%%%%%%%%%%%%%%%%%%%%%%%%%%%%%%%%%%%%%%%%%%%%%%%%%%%%%%%%%%%%%%%%%%%%%%%%%%%%%%%%%%
%%%%%%%%%%%%%%%%%%%%%%%%%%%%%%%%%%%%%%%%%%%%%%%%%%%%%%%%%%%%%%%%%%%%%%%%%%%%%%%%%%%%%%%%%%%%%%%%%%%%%%%%%%%%%%%%
Unlike decoding by user $\mathrm{B}$, for decoding by the adversary $\mathrm{E}$, the authors of~\cite{DG13v1} claimed that the complexity of a problem called in~\cite{DG13v1} the ``Decision'' variant of the ``\aln{MIMO} decoding problem'' (to be called \aln{MIMO-Decision} from here on), namely distinguishing between samples of two distribution $\mathcal{A}_{m,\alpha}$ and $\mathcal{R}_{\alpha}$ both defined on $\mathbb{R}^{n_t}\times\mathbb{R}$. The first one is the distribution of the channel coefficients and the received signal from a single antenna in a \aln{MIMO} channel. Since there are $n_r$ receive antenna, there will be $n_r$ samples of $\mathcal{A}_{m,\alpha}$. The second one is basically going to be identical to the first one lacking the underlying structure. The authors of~\cite{DG13v1} then claimed a security reduction to \aln{MIMO-Search} problem, that is recovering  ${\bf x}$ from ${\bf y}' = {\bf G}_v{\bf x} + {\bf e}'$ and ${\bf G}_v$, with non-negligible probability, under certain parameter settings, upon using massive \aln{MIMO} systems with large number of transmit antennas $n_t$. And finally they claimed that the \aln{MIMO-Search} is as hard as solving standard lattice problems in the worst-case. More precisely, it was claimed in~\cite{DG13v1} that, upon considering above conditions, user $\mathrm{E}$ will face an exponential complexity in decoding the message ${\bf x}$. For our security analysis, we focus here for simplicity on this \aln{MIMO-Search} variant. We say that the \aln{MIMO-Search} problem is \emph{hard} (and the \aln{MMPLC-13} is \emph{secure} in the sense of ``one-wayness'') if any attack algorithm against \aln{MIMO-Search} with run-time $\poly(n_t)$ has negligible success probability $n_t^{-\omega(1)}$.
More precisely, in Theorem 1 of~\cite{DG13v1}, a polynomial-time complexity reduction is claimed from worst-case instances of the $\aln{GapSVP}_{n_t/\alpha}$ problem in arbitrary lattices of dimension $n_t$, to the \aln{MIMO-Search} problem with $n_t$ transmit antennas, noise parameter $\alpha$ and constellation size $m$, assuming the following minimum noise level for the equivalent channel in between $\mathrm{A}$ and $\mathrm{E}$ holds:
\begin{equation}~\label{eq:const1}
m\alpha>\sqrt{n_t}.
\end{equation}
The reduction is quantum when $m = \poly(n_t)$ and classical when $m = \mathcal{O}(2^{n_t})$, and is claimed to hold for \emph{any polynomial number of receive antennas} $n'_r = \poly(n_t)$. We show in Section~\ref{Section:Zero-ForcingAttack}, however, that in fact for
$$m \alpha < c n'_r/\sqrt{\log n'_r},$$
(which does not violate \eqref{eq:const1}) for some constant $c$, there exists an efficient algorithm (Zero-Forcing linear receiver) for \aln{MIMO-Search}. Since \eqref{eq:const1} is independent of the number of receive antennas $n_r'$, the condition \eqref{eq:const1} turns out to be not sufficient to provide security of the \aln{MMPLC-13}. We will provide our detailed analysis in Section~\ref{Section:Zero-ForcingAttack}.

%%%%%%%%%%%%%%%%%%%%%%%%%%%%%%%%%%%%%%%%%%%%%%%%%%%%%%%%%%%%%%%%%%%%%%%%%%%%%%%%%%%%%%%%%%%%%%%%%%%%%%%%%%%%%%%%
%%%%%%%%%%%%%%%%%%%%%%%%%%%%%%%%%%%%%%%%%%%%%%%%%%%%%%%%%%%%%%%%%%%%%%%%%%%%%%%%%%%%%%%%%%%%%%%%%%%%%%%%%%%%%%%%
\subsection{Security Condition of the Cryptosystem in~\cite{DG13v2}}\label{SubSection:SecurityCondition2}
%%%%%%%%%%%%%%%%%%%%%%%%%%%%%%%%%%%%%%%%%%%%%%%%%%%%%%%%%%%%%%%%%%%%%%%%%%%%%%%%%%%%%%%%%%%%%%%%%%%%%%%%%%%%%%%%
%%%%%%%%%%%%%%%%%%%%%%%%%%%%%%%%%%%%%%%%%%%%%%%%%%%%%%%%%%%%%%%%%%%%%%%%%%%%%%%%%%%%%%%%%%%%%%%%%%%%%%%%%%%%%%%%
The security of the cryptosystem provided in~\cite{DG13v2} is claimed based on the hardness of \aln{MIMO-Search} problem, explained above. However, the hardness conditions are different from that of~\cite{DG13v1}. In Theorem 1 of~\cite{DG13v2}, a polynomial-time complexity reduction is claimed from worst-case instances of the $\aln{GapSVP}_{n_t/\alpha}$ and $\aln{SIVP}_{n_t/\alpha}$ problems in arbitrary lattices of dimension $n_t$, to the \aln{MIMO-Search} problem with $n_t$ transmit antennas, noise parameter $\alpha$ and constellation size $m$, assuming the following two hardness conditions hold:
\begin{equation}~\label{eq:const21}
m \geq n_r 2^{n_t \log \log n_t / \log n_t},
\end{equation}
and
\begin{equation}~\label{eq:const22}
n_r\alpha \geq 2\pi\sqrt{n_t}.
\end{equation}
Notice that the number of transmit antennas $n$, signal constellation $M$-PAM, and the number of receive antennas $m$ in~\cite{DG13v2} are simply replaced by our notations $n_t$, $m$-PAM, and $n_r$, respectively. Furthermore, the second condition \eqref{eq:const22} is originally $n_r\alpha/k^2 \geq \sqrt{n_t}$, where $k/\sqrt{2\pi}$ is the standard deviation of the entries of the channel gain matrix ${\bf H}$. However, without loss of generality and for simplicity, we assume $k=\sqrt{2\pi}$, which results in \eqref{eq:const22}.
%%%%%%%%%%%%%%%%%%%%%%%%%%%%%%%%%%%%%%%%%%%%%%%%%%%%%%%%%%%%%%%%%%%%%%%%%%%%%%%%%%%%%%%%%%%%%%%%%%%%%%%%%%%%%%%%
%%%%%%%%%%%%%%%%%%%%%%%%%%%%%%%%%%%%%%%%%%%%%%%%%%%%%%%%%%%%%%%%%%%%%%%%%%%%%%%%%%%%%%%%%%%%%%%%%%%%%%%%%%%%%%%%
\section{Zero-Forcing Attack on Cryptosystem in~\cite{DG13v1}}\label{Section:Zero-ForcingAttack}
%%%%%%%%%%%%%%%%%%%%%%%%%%%%%%%%%%%%%%%%%%%%%%%%%%%%%%%%%%%%%%%%%%%%%%%%%%%%%%%%%%%%%%%%%%%%%%%%%%%%%%%%%%%%%%%%
%%%%%%%%%%%%%%%%%%%%%%%%%%%%%%%%%%%%%%%%%%%%%%%%%%%%%%%%%%%%%%%%%%%%%%%%%%%%%%%%%%%%%%%%%%%%%%%%%%%%%%%%%%%%%%%%
In this section, we introduce a simple and efficient attack based on \aln{ZF} linear receivers~\cite{Kumar09} to the \aln{MMPLC-13} cryptosystem of~\cite{DG13v1}. In particular, we show that user $\mathrm{E}$ can employ an efficient algorithm (that is \aln{ZF} linear receiver) on its received signal and decode the plain-text ${\bf x}$ with overwhelming probability. Such an algorithm implies that \aln{MIMO-Search} problem is not hard as it is claimed in~\cite{DG13v1}. We first introduce the attack and analyze its components. The eavesdropper $\mathrm{E}$ receives ${\bf y}' = {\bf G}_v{\bf x} + {\bf e}'$. Let ${\bf G}_v = {\bf U}'{\bf \Sigma}'({\bf V}')^t$ be the \aln{SVD} of the equivalent channel ${\bf G}_v$. Thus, we get
${\bf y}' = {\bf U}'{\bf \Sigma}'({\bf V}')^t{\bf x} + {\bf e}'$,
where both ${\bf U}'$ and ${\bf V}'$ are orthogonal matrices and ${\bf \Sigma}'$ equals
$\mbox{diag}\left(\sigma_1({\bf G}_v),\ldots,\sigma_{n_t}({\bf G}_v)\right)=\mbox{diag}\left(\sigma_1({\bf G}),\ldots,\sigma_{n_t}({\bf G})\right)$,
where the last equality holds since the singular values of ${\bf G}_v$ and ${\bf G}$ are the same. Note that $\mathrm{E}$ knows ${\bf G}_v$ and its \aln{SVD} from the assumption that (s)he knows the channel between $\mathrm{A}$ and $\mathrm{B}$. At this point, user $\mathrm{E}$ performs a \aln{ZF} attack~\cite{Kumar09}. S(he) computes
\begin{equation}~\label{eq:Eveequation}
\tilde{\bf y}' = ({\bf G}_v)^{-1} {\bf y}' = {\bf x} + \tilde{\bf e}',
\end{equation}
where $\tilde{\bf e}' = ({\bf G}_v)^{-1}{\bf e}' = {\bf V}'({\bf \Sigma}')^{-1}({\bf U'})^t{\bf e}'$. User $\mathrm{E}$ is now able to recover an estimate $\tilde{x}'_i$ of the $i$-th coordinate $x_i$ of ${\bf x}$, by rounding:
$\tilde{x}'_i = \left\lceil \tilde{y}_i'\right\rfloor = \left\lceil x_i + \tilde{e}'_i\right\rfloor = x_i + \left\lceil\tilde{e}'_i\right\rfloor$.
Let ${\bf v}_i'=(v'_{i,1},\ldots,v'_{i,n_t})$ denotes the $i$-th row of ${\bf V}'$, we define
\begin{equation}~\label{eq:sigmant}
\sigma^2_{t_i}\triangleq m^2\alpha^2\sum_{j=1}^{n_t} \!\frac{|v'_{i,j}|^2}{\sigma^2_j({\bf G})}.
\end{equation}
%%%%%%%%%%%%%%%%%%%%%%%%%%%%%%%%%%%%%%%%%%%%%%%%%%%%%%%%%%%%%%%%%%%%%%%%%%%%%%%%%%%%%%%%%%%%%%%%%%%%%%%%%%%%%%%%
%%%%%%%%%%%%%%%%%%%%%%%%%%%%%%%%%%%%%%%%%%%%%%%%%%%%%%%%%%%%%%%%%%%%%%%%%%%%%%%%%%%%%%%%%%%%%%%%%%%%%%%%%%%%%%%%
\subsection{Analysis of \aln{ZF} Attack}\label{SubSection:NoiseAnalysis}
%%%%%%%%%%%%%%%%%%%%%%%%%%%%%%%%%%%%%%%%%%%%%%%%%%%%%%%%%%%%%%%%%%%%%%%%%%%%%%%%%%%%%%%%%%%%%%%%%%%%%%%%%%%%%%%%
%%%%%%%%%%%%%%%%%%%%%%%%%%%%%%%%%%%%%%%%%%%%%%%%%%%%%%%%%%%%%%%%%%%%%%%%%%%%%%%%%%%%%%%%%%%%%%%%%%%%%%%%%%%%%%%%
We now investigate the distribution of $\tilde{\bf e}'$ in \eqref{eq:Eveequation}.
\begin{lemma}~\label{lem:noiseatEVE}
The $i$-th component of $\tilde{\bf e}'$ in \eqref{eq:Eveequation} is distributed as $\mathcal{N}_{\sigma^2_{t_i}}$, for $1\leq i \leq n_t$, where $\sigma^2_{t_i}$ is defined in \eqref{eq:sigmant}. Furthermore, if $\sigma_{\mathrm{E}}^2=\max_i\{\sigma^2_{t_i}\}$, then
$$\sigma_{\mathrm{E}}^2\leq m^2\alpha^2/\sigma^2_{n_t}({\bf G}).$$
\end{lemma}
\begin{IEEEproof}
	See appendix~\ref{appen}.
\end{IEEEproof}
The above explained \aln{ZF} attack succeeds if $|\tilde{e}'_i| < 1/2$ for all $1\leq i \leq n_t$. Let $\mathbb{P}_{\tiny \aln{ZF}}\left[\mathrm{E}|{\bf G}\right]$ denotes the decoding error probability that $\mathrm{E}$ incorrectly recovers ${\bf x}$ using \aln{ZF} attack. Based on Lemma~\ref{lem:noiseatEVE}, we have
\begin{eqnarray}
\mathbb{P}_{\tiny \aln{ZF}}\left[\mathrm{E}|{\bf G}\right]&\leq&n_t\mathbb{P}_{w \samp \mathcal{N}_{\sigma^2_{\mathrm{E}}}}\left[|w| > 1/2\right]\nonumber\\
&\leq&n_t\mathbb{P}_{w \samp \mathcal{N}_{1}}\left[|w| > |\sigma_{n_t}({\bf G})|/(2 m \alpha) \right]\label{Pe:AE}\\
&\leq&n_t\exp\left(-|\sigma_{n_t}({\bf G})|^2/(8 m^2 \alpha^2) \right).~\label{Pe:AE1}
\end{eqnarray}
By comparing \eqref{Pe:AB} and \eqref{Pe:AE}, we see that the noise conditions for decoding ${\bf x}$ by users $\mathrm{B}$ and $\mathrm{E}$ are the same if both users have the same number of receive antennas $n_r'=n_r$ and the distributions of channels ${\bf G}$ and ${\bf H}$ are the same. This implies that user $\mathrm{E}$ is able to decode under the same constraints/conditions as $\mathrm{B}$. Moreover, if $n_r'> n_r$, then the adversary $\mathrm{E}$ is capable of decoding in the presence of stronger noise.
\begin{remark}
We show that in case of considering the hardness condition from~\cite{DG13v1}, the upper bound \eqref{Pe:AB1} on the error probability of legitimate decoder $\mathrm{B}$ is asymptotically equal to the upper bound \eqref{Pe:AE} on the error probability of the attacker decoder $\mathrm{E}$ if the latter uses a \aln{ZF} attack. One may object that the upper bound in \eqref{Pe:AB0} is not sharp; indeed, the union bound can be tightened to $$\sum_{i=1}^{n_t}\mathbb{P}_{w \samp \mathcal{N}_{m^2 \alpha^2}}\left[|w| > |\sigma_{i}({\bf H})|/2\right],$$
and in general, the exact probability of incorrectly decoding ${\bf x}$ by $\mathrm{B}$ is
$$1-\prod_i(1-\mathbb{P}_{w \samp \mathcal{N}_{m^2 \alpha^2}}\left[|w| > |\sigma_{i}({\bf H})|/2\right]),$$
which might be less than the value in the above summation and the one in \eqref{Pe:AB0}. However, this looseness of the bound does not significantly change our conclusions for the following reason. Notice that $\mathrm{B}$'s error probability is lower bounded as
\begin{equation}~\label{eq:lbPBH}
\mathbb{P}_{w \samp \mathcal{N}_{m^2 \alpha^2}}\left[|w| > |\sigma_{n_t}({\bf H})|/2\right]\leq\mathbb{P}\left[\mathrm{B}|{\bf H}\right].
\end{equation}
Comparing the lower bound \eqref{Pe:AB0} with the upper bound \eqref{eq:lbPBH} on $\mathbb{P}\left[\mathrm{B}|{\bf H}\right]$ shows that the latter exceeds the lower bound by at most a linear factor $n_t$. Therefore, even taking the looseness of the bound into account, if the parameters are chosen to make the legitimate decoder's error probability $\mathbb{P}\left[\mathrm{B}|{\bf H}\right] \leq 1/n_t^{\omega(1)}$ negligible (which is needed for the correctness of the system), then our results (Lemma~\ref{lem:noiseatEVE} and \eqref{Pe:AE}) show that attacker decoder's error probability $\mathbb{P}\left[\mathrm{E}|{\bf G}\right]$ is equivalent to $\mathbb{P}\left[\mathrm{B}|{\bf H}\right]$ and that is $\leq n_t \cdot 1/n_t^{\omega(1)} \leq 1/n_t^{\omega(1)}$ if $n_r'$ grows larger than $n_t$, and hence also negligible. %Secondly, as we will again study its asymptotic behaviour in comparison to an upper bound on the probability of error for user $\mathrm{E}$ in subsection~\ref{SubSection:GeneralPrecodingScheme}this upper bound becomes tight $\mathbb{P}\left[\mathrm{B}|{\bf H}\right]$ is lower bounded by $\exp\left(-|\sigma_{1}({\bf H})|^2/\left(8 m^2 \alpha^2\right)\right)$, . In particular, we show that a lower bound on $\mathbb{P}\left[\mathrm{B}|{\bf H}\right]$ is asymptotically weaker than an upper bound on $\mathbb{P}\left[\mathrm{E}|{\bf G}\right]$, that is the probability of incorrectly decoding the sent message by user $\mathrm{E}$ employing an specific attack (see Section~\ref{Section:Zero-ForcingAttack}).
\end{remark}
%%%%%%%%%%%%%%%%%%%%%%%%%%%%%%%%%%%%%%%%%%%%%%%%%%%%%%%%%%%%%%%%%%%%%%%%%%%%%%%%%%%%%%%%%%%%%%%%%%%%%%%%%%%%%%%%
%%%%%%%%%%%%%%%%%%%%%%%%%%%%%%%%%%%%%%%%%%%%%%%%%%%%%%%%%%%%%%%%%%%%%%%%%%%%%%%%%%%%%%%%%%%%%%%%%%%%%%%%%%%%%%%%
\subsection{Asymptotic Probability of Error for Adversary}\label{SubSection:AsymptoticProbabilityofErrorforAdversary}
%%%%%%%%%%%%%%%%%%%%%%%%%%%%%%%%%%%%%%%%%%%%%%%%%%%%%%%%%%%%%%%%%%%%%%%%%%%%%%%%%%%%%%%%%%%%%%%%%%%%%%%%%%%%%%%%
%%%%%%%%%%%%%%%%%%%%%%%%%%%%%%%%%%%%%%%%%%%%%%%%%%%%%%%%%%%%%%%%%%%%%%%%%%%%%%%%%%%%%%%%%%%%%%%%%%%%%%%%%%%%%%%%
Before starting this section, we mention a Theorem from~\cite{Edelman89} regarding the least/largest singular value of matrix variate Gaussian distribution. This theorem relates the least/largest singular value of a Gaussian matrix to the number of its columns and rows asymptotically.
\begin{theorem}[\cite{Edelman89}]~\label{th:sv}
Let ${\bf M}$ be an $s\times t$ matrix with i.i.d. entries distributed as $\mathcal{N}_1$. If $s$ and $t$ tend to infinity in such a way that $s/t$ tends to a limit $y\in[1,\infty]$, then
\begin{equation}~\label{eq:advratio}
\sigma^2_t({\bf M})/s\rightarrow \left(1-1/\sqrt{y}\right)^2
\end{equation}
and
\begin{equation}~\label{eq:upLsv}
\sigma^2_1({\bf M})/s\rightarrow \left(1+1/\sqrt{y}\right)^2,
\end{equation}
almost surely.
\end{theorem}
We now analyze the asymptotic probability of error for eavesdropper using a \aln{ZF} linear receiver.
\begin{theorem}~\label{th:PeZF}
Fix any real $\varepsilon,\varepsilon'>0$, and $y' \in [1,\infty]$, and suppose that $n_r'/n_t\rightarrow y'$ as $n_t \rightarrow \infty$. Then, for all sufficiently large $n_t$, the probability $\mathbb{P}_{\tiny \aln{ZF}}[\mathrm{E}]$ that $\mathrm{E}$ incorrectly decodes the message ${\bf x}$ using a \aln{ZF} decoder is upper bounded by $\varepsilon$, if
\begin{equation}~\label{eq:PeZF}
m^2\alpha^2\leq n_r'\left(\left(1-1/\sqrt{y'}\right)^2-\varepsilon'\right)/\left(8\log\left(2n_t/\varepsilon\right)\right).
\end{equation}
\end{theorem}
\begin{IEEEproof}
See Appendix~\ref{appen}.
\end{IEEEproof}
Comparing conditions \eqref{eq:const1} and \eqref{eq:PeZF}, we conclude that if $y'$ exceeds a small factor at most logarithmic in $n_t$, i.e. $y' = \mathcal{O}(\log n_t)$ we can have both conditions satisfied and yet Theorem~\ref{th:PeZF} shows that \aln{MIMO-Search} can be efficiently solved, i.e. this contradicts the hardness of the \aln{MIMO-Search} problem conjectured in~\cite{DG13v1} to hold for much larger polynomial ratios $y' = O(\poly(n_t))$.

To analytically investigate the advantage of decoding at $\mathrm{B}$ over $\mathrm{E}$, we define the following advantage ratio.
\begin{definition}~\label{definition:advrationZF}
	For fixed channel matrices ${\bf H}$ and ${\bf G}$, the ratio
\begin{equation}~\label{def:advrationZF}
	\mbox{adv}\triangleq \frac{\log\mathbb{P}(\mathrm{B}|{\bf H})}{\log\mathbb{P}(\mathrm{E}|{\bf G})},
\end{equation}
is called the advantage of $\mathrm{B}$ over $\mathrm{E}$.
\end{definition}
The above advantage ratio is suitable to capture the decoding advantage of user $\mathrm{B}$ over $\mathrm{E}$ asymptotically as it uses $\log$ function in its definition. In fact, it shows how faster the probability of error decays for user $\mathrm{B}$ than user $\mathrm{E}$. Note that such an advantage can be re-written for specific decoding algorithms too. For example, in the framework of \aln{MMPLC}, user $\mathrm{B}$ will always experience a diagonal channel and hence can decode ${\bf x}$ using a method explained in Subsection~\ref{SubSection:CorrectnessCondition}. If user $\mathrm{E}$ uses a \aln{ZF} linear receiver (as discussed so far), the advantage ratio with respect to \aln{ZF} attack is:
\begin{equation}~\label{def:advratio}
\mbox{adv}=\frac{\sigma^2_{n_t}({\bf H})}{\log\mathbb{P}_{\tiny \aln{ZF}}(\mathrm{E}|{\bf G})},
\end{equation}
which is upper bounded by
\begin{equation}~\label{def:advratio1}
\mbox{adv}_{\tiny \aln{ZF}}\triangleq \frac{\sigma^2_{n_t}({\bf H})}{\sigma^2_{n_t}({\bf G})},
\end{equation}
since \eqref{Pe:AB1} and \eqref{Pe:AE1} hold.
We note from \eqref{Pe:AB} and \eqref{Pe:AE} that $\mbox{adv}_{\tiny \aln{ZF}}$ is the ratio between the maximum noise power tolerated by $\mathrm{B}$'s \aln{ZF} decoder to the maximum noise power tolerated by $\mathrm{E}$'s \aln{ZF} decoder, for the same decoding error probability in both cases. First, we study this advantage ratio asymptotically. We use Theorem~\ref{th:sv} and substitute the obtained limits into \eqref{def:advratio} to get the following result.
\begin{proposition}~\label{prop:rectangularMMPLCdisadv}
Let ${\bf H}_{n_r \times n_t}$ be the channel between $\mathrm{A}$ and $\mathrm{B}$ and ${\bf G}_{n_r' \times n_t}$ be the channel between $\mathrm{A}$ and $\mathrm{E}$, both with i.i.d. elements each with distribution $\mathcal{N}_1$. Fix real $y,y' \in [1,\infty]$, and suppose that $n_r/n_t\rightarrow y$ and $n_r'/n_t\rightarrow y'$ as $n_t \rightarrow \infty$. Then, using a \aln{SVD} precoding technique in \aln{MM-PLC}, we have
$$\mbox{adv}_{\tiny \aln{ZF}} \rightarrow \left(\sqrt{y}-1\right)^2/\left(\sqrt{y'}-1\right)^2$$
almost surely as $n_t \rightarrow \infty$.
\end{proposition}
%\begin{IEEEproof}
%Based on Theorem~\ref{th:sv} for ${\bf H}$ and ${\bf G}$, we have
%$$\left\{\begin{array}{l}
%\sigma^2_{n_t}({\bf H})/n_r\rightarrow \left(1-1/\sqrt{y}\right)^2\\
%\sigma^2_{n_t}({\bf G})/n_r'\rightarrow \left(1-1/\sqrt{y'}\right)^2.
%\end{array}\right.$$
%Substituting the above two limits into \eqref{def:advratio} and using $n_r/n_r' = (n_r/n_t)/(n'_r/n_t) \rightarrow %y/y'$, the result follows.
%\end{IEEEproof}
Note that $\mbox{adv}_{\tiny \aln{ZF}}\rightarrow1$ is obtained in the case that $y=y'$, which is equivalent to $n_r/n_r'\rightarrow1$. On the other hand $\mbox{adv}_{\tiny \aln{ZF}}\rightarrow0$, if $y'/y = \infty$ which is equivalent to $n_r'/ n_r \rightarrow \infty$.

\begin{remark}
Note that the above defined advantage ratio captures an analytical attack on \aln{MIMO-Search} problem and consequently \aln{MMPLC-13}. For numerical results/analysis of such an attack, we refer the reader to~\cite{SS15c} and~\cite{attack15}.
\end{remark}
%%%%%%%%%%%%%%%%%%%%%%%%%%%%%%%%%%%%%%%%%%%%%%%%%%%%%%%%%%%%%%%%%%%%%%%%%%%%%%%%%%%%%%%%%%%%%%%%%%%%%%%%%%%%%%%%
%%%%%%%%%%%%%%%%%%%%%%%%%%%%%%%%%%%%%%%%%%%%%%%%%%%%%%%%%%%%%%%%%%%%%%%%%%%%%%%%%%%%%%%%%%%%%%%%%%%%%%%%%%%%%%%%
\subsection{General Precoding Scheme}\label{SubSection:GeneralPrecodingScheme}
%%%%%%%%%%%%%%%%%%%%%%%%%%%%%%%%%%%%%%%%%%%%%%%%%%%%%%%%%%%%%%%%%%%%%%%%%%%%%%%%%%%%%%%%%%%%%%%%%%%%%%%%%%%%%%%%
%%%%%%%%%%%%%%%%%%%%%%%%%%%%%%%%%%%%%%%%%%%%%%%%%%%%%%%%%%%%%%%%%%%%%%%%%%%%%%%%%%%%%%%%%%%%%%%%%%%%%%%%%%%%%%%%
One may wonder whether a different precoding method (again, assumed known to $\mathrm{E}$) than used above may provide a better advantage ratio for $\mathrm{B}$ over $\mathrm{E}$. %However, this also seems to basically fail.
Suppose that instead of sending $\tilde{\bf x} = {\bf V}{\bf x}$, user $\mathrm{A}$ precodes $\tilde{\bf x} = {\bf P}({\bf H}){\bf x}$, where ${\bf P} = {\bf P}({\bf H})$ is some other precoding matrix that depends on the channel matrix ${\bf H}$. Then, given the channel matrices, the analysis given in Section~\ref{Section:Zero-ForcingAttack} shows that using \aln{ZF} decoding, $\mathrm{B}$'s decoding error probability will be bounded as
$$ n_t \exp\left(\left(-\sigma^2_{n_t}({\bf H}{\bf P})\right)/\left(8m^2\alpha^2\right)\right),$$
while $\mathrm{E}$'s decoding error probability will be bounded as
$$ n_t \exp\left(\left(-\sigma^2_{n_t}({\bf G}{\bf P})\right)/\left(8m^2\alpha^2\right)\right).$$
Therefore, in this general case, the advantage ratio of maximum noise power decodable by $\mathrm{B}$ to that decodable by $\mathrm{E}$ under a \aln{ZF} attack at a given error probability generalizes from \eqref{def:advratio} to
\begin{equation}~\label{def:advratiogen}
\mbox{adv}_{\tiny \aln{ZF}}\triangleq \sigma^2_{n_t}({\bf H}{\bf P})/\sigma^2_{n_t}({\bf G}{\bf P}).
\end{equation}
We now give an upper bound on the advantage ratio \eqref{def:advratiogen}. Let us first define
$$\mbox{advup}_{\tiny \aln{ZF}}\triangleq\sigma^2_1({\bf H})/\sigma^2_{n_t}({\bf G}).$$
\begin{proposition}~\label{prop:rectangularMMPLCGeneralPrecoderdisadv}
Let ${\bf H}$ and ${\bf G}$ be as in Proposition~\ref{prop:rectangularMMPLCdisadv}. Then we have $\mbox{adv}_{\tiny \aln{ZF}}\leq\mbox{advup}_{\tiny \aln{ZF}}$. Furthermore, fix real $y,y' \in [1,\infty]$, and suppose that $n_r/n_t\rightarrow y$ and $n_r'/n_t\rightarrow y'$ as $n_t \rightarrow \infty$, so that $n_r'/n_r \rightarrow y'/y \triangleq \rho'$. Then, using a general precoding matrix ${\bf P}({\bf H})$ in \aln{MM-PLC}, we have
$$\mbox{advup}_{\tiny \aln{ZF}} \rightarrow \left(\sqrt{y}+1\right)^2/\left(\sqrt{y'}-1\right)^2$$
almost surely as $n_t \rightarrow \infty$. Hence, in the case $n'_r=n_r$ and $y'=y \rightarrow \infty$, we have $\mbox{advup}_{\tiny \aln{ZF}} \rightarrow 1$. Moreover, if $\mbox{advup}_{\tiny \aln{ZF}} \rightarrow c$ for some $c \geq 1$, then $\min(y',\rho') \leq 9$. %using $n_r'=n_r$ and $y=y'\rightarrow \infty$, which are both within a polynomial of $n_t$.
\end{proposition}
\begin{IEEEproof}
See Appendix~\ref{appen}.
\end{IEEEproof}
\begin{remark}
Notice that to derive the results of Proposition~\ref{prop:rectangularMMPLCdisadv} and~\ref{prop:rectangularMMPLCGeneralPrecoderdisadv} we have used the randomness of channels ${\bf H}$ and ${\bf G}$. It means that our results show that $\mbox{adv}_{\aln{ZF}} \rightarrow 1$ and $\mbox{advup}_{\tiny \aln{ZF}} \rightarrow 1$ for the average-case in cryptographic senses (for a definition see~\cite{KatzLindell2007}). This simply implies that our analysis are also valid and might get stronger for the worst-case scenario, as worst-case is always worse that the average-case. 
\end{remark}
%%%%%%%%%%%%%%%%%%%%%%%%%%%%%%%%%%%%%%%%%%%%%%%%%%%%%%%%%%%%%%%%%%%%%%%%%%%%%%%%%%%%%%%%%%%%%%%%%%%%%%%%%%%%%%%%
%%%%%%%%%%%%%%%%%%%%%%%%%%%%%%%%%%%%%%%%%%%%%%%%%%%%%%%%%%%%%%%%%%%%%%%%%%%%%%%%%%%%%%%%%%%%%%%%%%%%%%%%%%%%%%%%
\section{The cryptosystem in~\cite{DG13v2} is incorrect}\label{SubSection:Discussion}
%%%%%%%%%%%%%%%%%%%%%%%%%%%%%%%%%%%%%%%%%%%%%%%%%%%%%%%%%%%%%%%%%%%%%%%%%%%%%%%%%%%%%%%%%%%%%%%%%%%%%%%%%%%%%%%%
%%%%%%%%%%%%%%%%%%%%%%%%%%%%%%%%%%%%%%%%%%%%%%%%%%%%%%%%%%%%%%%%%%%%%%%%%%%%%%%%%%%%%%%%%%%%%%%%%%%%%%%%%%%%%%%%
We first note that the updated \aln{MMPLC-17} is that the basic system model is still the same, only the parameter choice (hardness conditions) for noise magnitude and constellation size has changed. Consequently, our analysis, which applies to the general model, for any choice of parameters, still applies. In particular, it still shows that user $\mathrm{B}$ has essentially no \aln{ZF} \aln{MIMO} decoding advantage over user $\mathrm{E}$ when user $\mathrm{E}$ has the same (or bigger) number of receiving antennas. Furthermore, we next show that the new larger noise/constellation parameter only makes unique \aln{MIMO} decoding by either user $\mathrm{E}$ or $\mathrm{B}$ information-theoretically impossible (not just computationally intractable), thus the \aln{MIMO} cryptosystem design \aln{MMPLC-17} is cryptographically incorrect. In particular, we show that user $\mathrm{B}$ cannot uniquely decode a sent message from $\mathrm{A}$, due to the large noise level imposed by the design to ensure security.

The updated \aln{MMPLC-17} in~\cite{DG13v2} works similar to that of~\cite{DG13v1}. However, to ensure security, the number of transmit antennas $n_t$, constellation size $m$, and the number of receive antennas $n_r$ should satisfy the following constraints:
\begin{equation}~\label{eq:const1v2}
m \geq n_r 2^{n_t \log \log n_t / \log n_t},
\end{equation}
and the modified noise condition is
\begin{equation}~\label{eq:const2v2}
n_r\alpha \geq 2\pi \sqrt{n_t}.
\end{equation}
Note that in~\cite{DG13v2}, the constellation size is denoted by $M$ and $m$ represents the number of receive antennas $n_r$. The latter is chosen by a user or a system to trade-off the noise requirement for constellation size $m$. If the noise level is below a certain threshold, efficient decoding methods such as \aln{ZF} linear receiver can attack the system again. This was studies at length in previous section. Since the constellation size $m$ is directly related to the decoding complexity and the security of the system, if the above conditions are not met, the results of~\cite{DG13v2} cannot provide any insight on the claimed security of \aln{MMPLC-17}.
We now give two results, which will prove useful in our analysis of \aln{MMPLC-17}. The first one is the Minkowski's First Theorem~\cite{ConwaySloane}.
\begin{theorem}~\label{th:minkowskisfirst}
Let $\Lambda_{\bf H}$ be a lattice generated by columns of ${\bf H}_{n_r\times n_t}$, then $\lambda_1(\Lambda)=\mathcal{O}(\sqrt{n_t})\det(\Lambda_{\bf H})^{1/n_t}$.
\end{theorem}
The second result finds an upper bound on the statistical distance (total variational distance) between a Gaussian distribution and its shifted one. The proof of the following result can be easily found by combining equations (8) and (10) of~\cite{ErHar12}.
\begin{lemma}~\label{lem:statisticaldistance}
Let $\mathcal{N}_{s^2}$ be a Gaussian distribution with zero mean and standard deviation $s$ and $\mathcal{N}_{s^2,\gamma}$ be $\gamma+\mathcal{N}_{s^2}$ (a shift of all samples of $\mathcal{N}_{s^2}$ by a constant $\gamma$), then
$$\Delta(\mathcal{N}_{s^2},\gamma+\mathcal{N}_{s^2})=\mathcal{O}(\gamma^2/s^2).$$
\end{lemma}
We now show that user $\mathrm{B}$ cannot uniquely decode the plain-text message ${\bf x}$ from its received signal considering the hardness conditions imposed to ensure security.
At one hand, we multiply both sides of~\eqref{eq:const1v2} by $\alpha$ and then combine the obtained inequality with the second condition \eqref{eq:const2v2}. It yields:
\begin{equation}~\label{eq:combinedv2}
m\alpha  \geq 2\pi 2^{n_t \log \log n_t / \log n_t}\sqrt{n_t}.
\end{equation}
On the other hand and based on Theorem~\ref{th:minkowskisfirst}, the approximate minimum distance of the lattices generated by ${\bf G}$ or ${\bf H}$ are in the order of $\mathcal{O}(\sqrt{n_t})$ with overwhelming probability when $n_r=\poly(n_t)$. Combining the above two arguments imply that, the noise standard deviation $m\alpha$ is sub-exponentially larger, by the factor $\eta(n_t) = 2^{n_t \log \log n_t / \log n_t}$, than the approximate minimum distance of both lattices ${\bf G}$ and ${\bf H}$. Note that this is not in contrast with neither \eqref{eq:const1v2} nor \eqref{eq:const2v2}. Therefore, both the legitimate user $\mathrm{B}$ and the adversary $\mathrm{E}$ are now in trouble decoding the plain-text message ${\bf x}$, since the received signal will fall outside a correct decoding sphere (centered at a lattice point with radius $\max\{\lambda_1(\Lambda_{\bf H})/2,\lambda_1(\Lambda_{\bf G})/2\}$) with high probability.
\begin{table*}[ht]
\centering
\caption{Summary of Our Results in Sections~\ref{Section:Zero-ForcingAttack}-\ref{SubSection:Discussion}. }~\label{table:summary}
\tabcolsep 0.02in
\begin{tabular}{||c|c|c|c|c||}
	\hline\hline
	Reference & Correctness Condition & Hardness Condition(s) & Hard Problem & Attack \\
	\hline
	\aln{MMPLC-13} in~\cite{DG13v1} & $|\tilde{e}_i| < \frac{|\sigma_i({\bf H})|}{2}, \forall1\leq i\leq n_t$ & $m\alpha>\sqrt{n_t}$ & \aln{MIMO}-Decision & \aln{ZF} attack (Section~\ref{Section:Zero-ForcingAttack})\\
	\hline
	\aln{MMPLC-17} in~\cite{DG13v2} & $|\tilde{e}_i| < \frac{|\sigma_i({\bf H})|}{2}, \forall 1\leq i\leq n_t$ & \begin{tabular}[x]{@{}c@{}}$n_r\alpha>\sqrt{n_t}$ and\\
		$m\geq n_r2^{n_t\log \log n_t/\log n_t}$\end{tabular} & \aln{MIMO}-Search & Correctness issue (Section~\ref{SubSection:Discussion})\\
	\hline\hline
\end{tabular}
\end{table*}
In particular, the following result is outstanding:
\begin{proposition}
For any fixed $\bf H$ and $\bf x$, the statistical distance $\Delta$ between ${\bf H}{\bf x}+{\bf e}$ and ${\bf H}({\bf x}+{\bf v}_1)+{\bf e}$ for ${\bf e}$ i.i.d. Gaussian with standard deviation $m\alpha$, and ${\bf v}_1=(1,0,\ldots,0)$ is sub-exponentially negligible.
\end{proposition}
\begin{IEEEproof}
It is obvious that the $\Delta({\bf H}{\bf x}+{\bf e}, {\bf H}({\bf x}+{\bf v}_1)+{\bf e})$ is less than or equal to the \aln{SD} between ${\bf H}{\bf v}_1+{\bf e}$ and ${\bf e}$, because of the common term ${\bf H}{\bf x}$. The latter itself is less than or equal to $n_r\Delta(e_1, \gamma + e_1)$ where $e_1$ is an $1$-dimensional Gaussian (because the $n_r$ components of ${\bf e}$ are independent), where $\gamma$ is an upper bound on the components of ${\bf H}{\bf v}_1$. In fact $\gamma= \mathcal{O}(\log n_r)$ with high probability, since the $n_r$ components in each column of ${\bf H}$ have standard deviation $\mathcal{O}(1)$. Now, based on Lemma~\ref{lem:statisticaldistance}, the statistical distance $\Delta(e_1, \gamma + e_1)$ between a Gaussian with standard deviation $m\alpha$ and its shift by $\gamma$ is $\mathcal{O}(\gamma^2/(m\alpha)^2)$. Consequently, for $n_r = \mathcal{O}(\poly(n_t))$, we have that $\Delta = \mathcal{O}(n_r \log (n_r)/\eta(n_t)) = 1/2^{\Omega(n_t/ \log n_t)}$ is sub-exponentially negligible for this scenario. Hence, even the legitimate user $\mathrm{B}$ cannot uniquely decode ${\bf x}$ under these new conditions.
\end{IEEEproof}
Since the statistical distance between ${\bf H}{\bf x}+{\bf e}$ and ${\bf H}({\bf x}+{\bf v}_1)+{\bf e}$ is sub-exponentially small, the legitimate user may decode either to ${\bf x}$ or ${\bf x}+{\bf v}_1$. Same ambiguity in decoding raises for ${\bf v}_j=(0,\ldots,0,1,0,\ldots,0)$, where there is a single $1$ at the $j$-th position and $0$ elsewhere, and therefore user $\mathrm{B}$ can decode to either ${\bf x}$ or ${\bf x}+{\bf v}_j$, $1\leq j\leq n_t$.
%%%%%%%%%%%%%%%%%%%%%%%%%%%%%%%%%%%%%%%%%%%%%%%%%%%%%%%%%%%%%%%%%%%%%%%%%%%%%%%%%%%%%%%%%%%%%%%%%%%%%%%%%%%%%%%%
%%%%%%%%%%%%%%%%%%%%%%%%%%%%%%%%%%%%%%%%%%%%%%%%%%%%%%%%%%%%%%%%%%%%%%%%%%%%%%%%%%%%%%%%%%%%%%%%%%%%%%%%%%%%%%%%
\section{Discussion on Potential of \aln{MMPLC}}~\label{AnUpperBoundonAdvantageRatio}
%%%%%%%%%%%%%%%%%%%%%%%%%%%%%%%%%%%%%%%%%%%%%%%%%%%%%%%%%%%%%%%%%%%%%%%%%%%%%%%%%%%%%%%%%%%%%%%%%%%%%%%%%%%%%%%%
%%%%%%%%%%%%%%%%%%%%%%%%%%%%%%%%%%%%%%%%%%%%%%%%%%%%%%%%%%%%%%%%%%%%%%%%%%%%%%%%%%%%%%%%%%%%%%%%%%%%%%%%%%%%%%%%
The results of the previous sections on both \aln{MMPLC-13} and \aln{MMPLC-17} are summarized in Table~\ref{table:summary}. It is now obvious from this Table that both \aln{MMPLC-13} and \aln{MMPLC-17} have some issues associated to them, the first one has got security issues, while the second one (the updated one) does not seem to have the security problem but cannot deliver a unique message to the legitimate user. However, we still see potential in \aln{MMPLC} approach. We discuss/discover in more details some properties of \aln{MMPLC} by changing some design criterion.

The analysis of Section~\ref{Section:Zero-ForcingAttack} shows that one cannot hope to achieve an advantage ratio greater than $1$, if the adversary uses a number of antennas significantly larger than used by the legitimate parties (by more than a constant factor). We now explore what advantage ratio can be achieved if we add a new constraint to \aln{MMPLC}, namely the number of adversary antennas is limited to be the same as the number of legitimate transmit and receive antennas. That is, we study the advantage ratio when the channel matrices ${\bf H}$ and ${\bf G}$ are square matrices and not rectangular. We show that under this simple constraint $n=n_t=n_r=n_r'$, the advantage ratio can get larger than $1$ and as big as $\mathcal{O}\left(n^2\right)$. We employ the following result in our analysis.
\begin{theorem}[Th. 5.1, \cite{Edelman89}]~\label{th:squarelsv}
Let ${\bf M}$ be a $t\times t$ matrix with i.i.d. entries distributed as $\mathcal{N}_1$. The least singular value of ${\bf M}$ satisfies
\begin{equation}~\label{eq:squarelsv}
\lim_{t\rightarrow\infty}\mathbb{P}\left[\sqrt{t}\sigma_t({\bf M})\geq x\right] = \frac{(1+x)\exp\left(-x^2/2-x\right)}{2x}.
\end{equation}
\end{theorem}
We note that for a similar result on the largest singular value for square matrices, Theorem~\ref{th:sv} is enough. Using the above Theorem along with Theorem~\ref{th:sv}, one can further upper bound and estimate the advantage ratio. More precisely, we have
\begin{eqnarray}
\mbox{adv}_{\tiny \aln{ZF}} &\leq& \sigma^2_1({\bf H})/\sigma^2_{n}({\bf G})\label{eq:in-1}\\
&\rightarrow &4n/\sigma^2_{n}({\bf G})=4n^2/\left(n\sigma^2_{n}({\bf G})\right),\label{eq:in0}
\end{eqnarray}
where \eqref{eq:in-1} is obtained based on \eqref{eq:advratioGeneralPrecoderMM-PLC}. As $n\rightarrow\infty$, based on Theorem~\ref{th:squarelsv}, the denominator of the RHS of \eqref{eq:in0} is $\mathcal{O}(1)$ except with probability $\leq \varepsilon$ for any fixed $\varepsilon>0$, and thus $\mbox{adv}_{\tiny \aln{ZF}}$ is $\mathcal{O}\left(n^2\right)$ with the same probability. The following proposition is now outstanding.
\begin{proposition}~\label{prop:squareub}
Let $\varepsilon>0$ be fixed, ${\bf H}$ and ${\bf G}$ be $n\times n$ matrices as in Proposition~\ref{prop:rectangularMMPLCdisadv} with $n=n_t=n_r=n_r'$. Using a general precoder ${\bf P}({\bf H})$ to send  ${\bf x}$, the maximum possible $\mbox{adv}_{\tiny \aln{ZF}}$ that $\mathrm{B}$ can achieve over $\mathrm{E}$, is of order $\mathcal{O}\left(n^2\right)$, except with probability $\leq \varepsilon$.
\end{proposition}
The above proposition implies that user $\mathrm{B}$ \emph{may} be able to decode the message ${\bf x}$, with noise power up to $n^2$ times greater than $\mathrm{E}$ is able to handle. Such an advantage was not available in \aln{MMPLC} scheme proposed in~\cite{DG13v1} due to the lack of constraint on the number of receive antennas for $\mathrm{E}$ and the use of \aln{SVD} precoder.

In the following, we present the achievability of results in Proposition~\ref{prop:squareub}, {i.e.} we show that the \aln{MMPLC} technique can approach the maximum achievable $\mbox{adv}_{\tiny \aln{ZF}}$ of order $\mathcal{O}\left(n^2\right)$ with $n_t=n_r=n_r'$ and an inverse precoder. This inverse precoder is definitely not  power efficient as it needs huge power enhancement at the transmitter, however it gives us a benchmark on the achievable advantage ratio. Notice that such a precoder would not be practical at all in the sense of Telecommunication theory, however it proves useful in theoretical sense as it shows that the upper bound on  advantage ratio is in fact sharp.

Throughout the rest of this section we assume two constraints 
\begin{equation}~\label{eq:InvPreCon}
n_t=n_r=n_r'~~~~\mbox{and}~~~~{\bf P}({\bf H})={\bf H}^{-1}.
\end{equation}
The equivalent channel between legitimate users is the identity matrix and the channel between users $\mathrm{A}$ and $\mathrm{E}$ is ${\bf G}{\bf H}^{-1}$. Thus, we have
$$\left\{\begin{array}{l}
\tilde{\bf y} = {\bf I}_n{\bf x} + \tilde{\bf e},\\
{\bf y}' = {\bf G}{\bf H}^{-1}{\bf x} + {\bf e}',
\end{array}\right.$$
Note that, for this framework the advantage ratio \eqref{def:advratio} under \aln{ZF} decoding algorithm at user $\mathrm{E}$ can be written as $1/\sigma_n^2\left({\bf G}{\bf H}^{-1}\right)$. We now proceed to find the distribution of $\sigma_n^2\left({\bf G}{\bf H}^{-1}\right)$, when both ${\bf G}$ and ${\bf H}$ are square standard Gaussian matrices of dimension $n$.

We say that a random variable $x$ has a distribution proportional to function $D$, if the $x$ is distributed as $c_0D $ for a constant $c_0$. We first find the distribution of ${\bf G}{\bf H}^{-1}$.
\begin{theorem}~\label{th:qutionet}
Let ${\bf Q} = {\bf G}{\bf H}^{-1}$, where ${\bf H}$ and ${\bf G}$ are two $n\times n$ real Gaussian matrices. 
\begin{itemize}
\item{} The distribution of ${\bf Q}$ is proportional to
\begin{equation}\label{eq:denQ}
1/\det\left({\bf I}_n+{\bf Q}{\bf Q}^t\right)^{n}.
\end{equation}
\item{} The joint probability density function of the eigenvalues of the product matrix ${\bf W}={\bf Q}{\bf Q}^t$ is proportional to
\begin{equation}~\label{eq:cv}
\prod_{\ell=1}^nw_{\ell}^{-\frac{1}{2}}\left(1-w_\ell\right)^{-\frac{1}{2}}\prod_{1\leq j < k \leq n}(w_k-w_j),
\end{equation} 
\end{itemize}
\end{theorem}
\begin{IEEEproof}
See Appendix~\ref{appen}.
\end{IEEEproof}
We now state the Selberg integral $s_n(\lambda_1,\lambda_2,\lambda)$ from~\cite{Forrester08,Forrester10}, which is defined as
$$\int_0^1\cdots\int_0^1\prod_{\ell=1}^nw_{\ell}^{\lambda_1}(1-w_{\ell})^{\lambda_2}\!\!\!\prod_{1\leq j < k \leq n}\!\!\!(w_k-w_j)^{2\lambda} dw_1 \cdots dw_n$$
and equals
\begin{equation}~\label{eq:SelbergInt}
\prod_{j=1}^{n-1}\frac{\Gamma(\lambda_1+1+j\lambda)\Gamma(\lambda_2+1+j\lambda)\Gamma(1+(1+j)\lambda)}{\Gamma(\lambda_1+\lambda_2+2+(n+j-1)\lambda)\Gamma(1+\lambda)},
\end{equation}
where $\Gamma(x)$ denotes the Gamma function.

The following theorem shows that using the setting of \eqref{eq:InvPreCon}, the decoding advantage $\mbox{adv}_{\tiny \aln{ZF}}$ of legitimate user $\mathrm{B}$ over adversary $\mathrm{E}$ with respect to \aln{ZF} attack approaches within a constant factor the upper bound $O(n^2)$ on $\mbox{adv}_{\tiny \aln{ZF}}$ from Section~\ref{SubSection:GeneralPrecodingScheme}, with probability arbitrarily close to $1$.
\begin{theorem}~\label{th:inverseprecoderachievability}
Let $\varepsilon>0$ be fixed, ${\bf H}$ and ${\bf G}$ be $n\times n$ Gaussian matrices as in Proposition~\ref{prop:rectangularMMPLCdisadv} with $n=n_t=n_r=n_r'$. Using an inverse precoder ${\bf P}({\bf H}) = {\bf H}^{-1}$ to send ${\bf x}$, the decoding advantage with respect to zero-forcing attack $\mbox{adv}_{\tiny \aln{ZF}}$, is at least $\frac{1}{4 \log(1/\varepsilon)} \cdot \left(n^2+n\right) = \Omega\left(n^2\right)$, except with probability $\leq \varepsilon$, for sufficiently large $n$.
\end{theorem}
\begin{IEEEproof}
See Appendix~\ref{appen}.
\end{IEEEproof}
An astute reader now asks why using \aln{ZF} linear receiver anymore, whereas there are more powerful \aln{MIMO} decoding algorithms including successive interference cancellation (\aln{SIC})~\cite{Babai86} and maximum likelihood (\aln{ML}) decoders~\cite{ViB}? In the next subsection, we address this question and show that using the setting of \eqref{eq:InvPreCon}, the advantage ratio with respect to \aln{SIC} is again non-trivial (in particular approaches $\mathcal{O}(n)$). We further show that since $n$ is chosen to be the security parameter in \aln{MMPLC}, it is essentially not practical to employ high-complex algorithms such as~\cite{ViB} neither for legitimate user nor for the adversary.
%%%%%%%%%%%%%%%%%%%%%%%%%%%%%%%%%%%%%%%%%%%%%%%%%%%%%%%%%%%%%%%%%%%%%%%%%%%%%%%%%%%%%%%%%%%%%%%%%%%%%%%%%%%%%%%%
\subsection{Adversary with \aln{SIC}}~\label{AdversarywithSIC}
%%%%%%%%%%%%%%%%%%%%%%%%%%%%%%%%%%%%%%%%%%%%%%%%%%%%%%%%%%%%%%%%%%%%%%%%%%%%%%%%%%%%%%%%%%%%%%%%%%%%%%%%%%%%%%%%
We now consider that user $\mathrm{E}$ performs successive interference cancellation (\aln{SIC})~\cite{Babai86}.
Let us also assume that ${\bf G}{\bf H}^{-1} = {\bf Q} = {\bf O} {\bf R}$ be the \aln{QR} decomposition of the equivalent channel, for an orthogonal matrix ${\bf O}$ and an upper triangular matrix ${\bf R}$ with diagonal elements $r_{jj}$, for $1\leq j \leq n$. Then the received vector by user $\mathrm{E}$ equals ${\bf y}' = {\bf O}{\bf R}{\bf x} + {\bf e}'$. Upon receiving ${\bf y}'$, this user multiplies it by ${\bf O}^t$, to obtain ${\bf y}''={\bf O}^t{\bf y}' = {\bf R}{\bf x} + {\bf O}^t{\bf e}'$. Hence, we get
$$\left\{\begin{array}{l}
\tilde{\bf y} = {\bf I}_n{\bf x} + \tilde{\bf e},\\
{\bf y}'' = {\bf R}{\bf x} + {\bf O}^t{\bf e}' = {\bf R}{\bf x} + {\bf e}'',
\end{array}\right.$$
In \aln{SIC} decoding framework, the last symbol is decoded first, {\em i.e.}
$$\tilde{x}'_n = \left\lfloor y''_n/r_{nn}\right\rceil=x_n+\left\lfloor e''_n/r_{nn}\right\rceil$$
is an estimate for $x_n$. The other symbols are approximated iteratively using
$$\tilde{x}'_j = \left\lfloor\frac{y''_j-\sum_{k=j+1}^nr_{jk}\tilde{x}'_k}{r_{jj}}\right\rceil,$$
for $j$ from $n-1$ downward to $1$. The error performance of such a decoder depends on the components of the diagonal entries of the ${\bf R}$ matrix. In other words, the above mentioned \aln{SIC} finds the closest vector if the distance from input vector to the lattice is less than half the length of $r_{nn}^2/2$.

\begin{figure}[htb]%
	\begin{center}%
		\includegraphics[width=7cm]{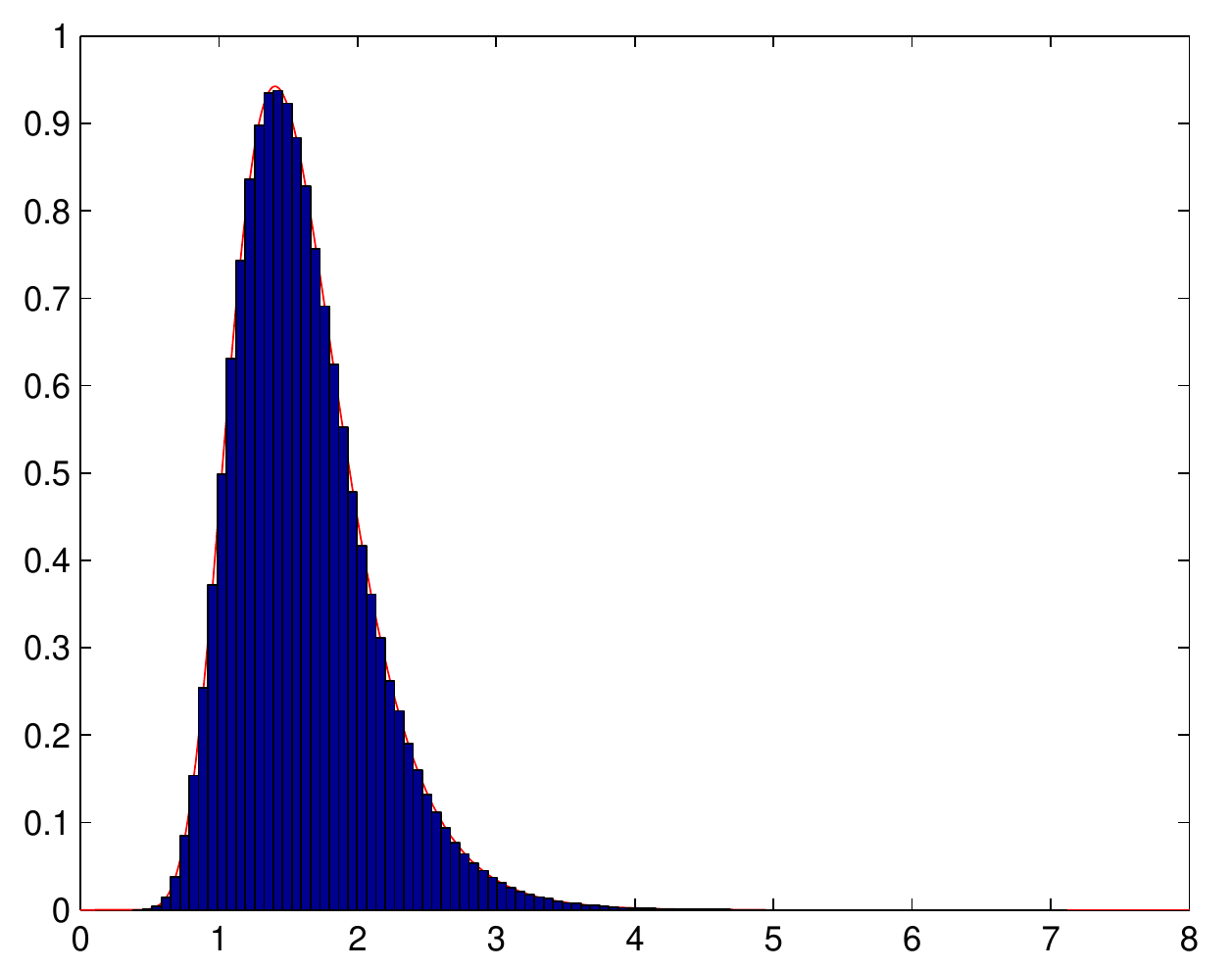}~\caption{\label{fig:histinvprec40} The histogram and the theoretical p.d.f. of $r_{jj}^2$ (red line) for $j=40$ and $10^6$ square channels of size $n=100$ using inverse precoder.}
	\end{center}
\end{figure}

\begin{figure}[htb]%
	\begin{center}%
		\includegraphics[width=7cm]{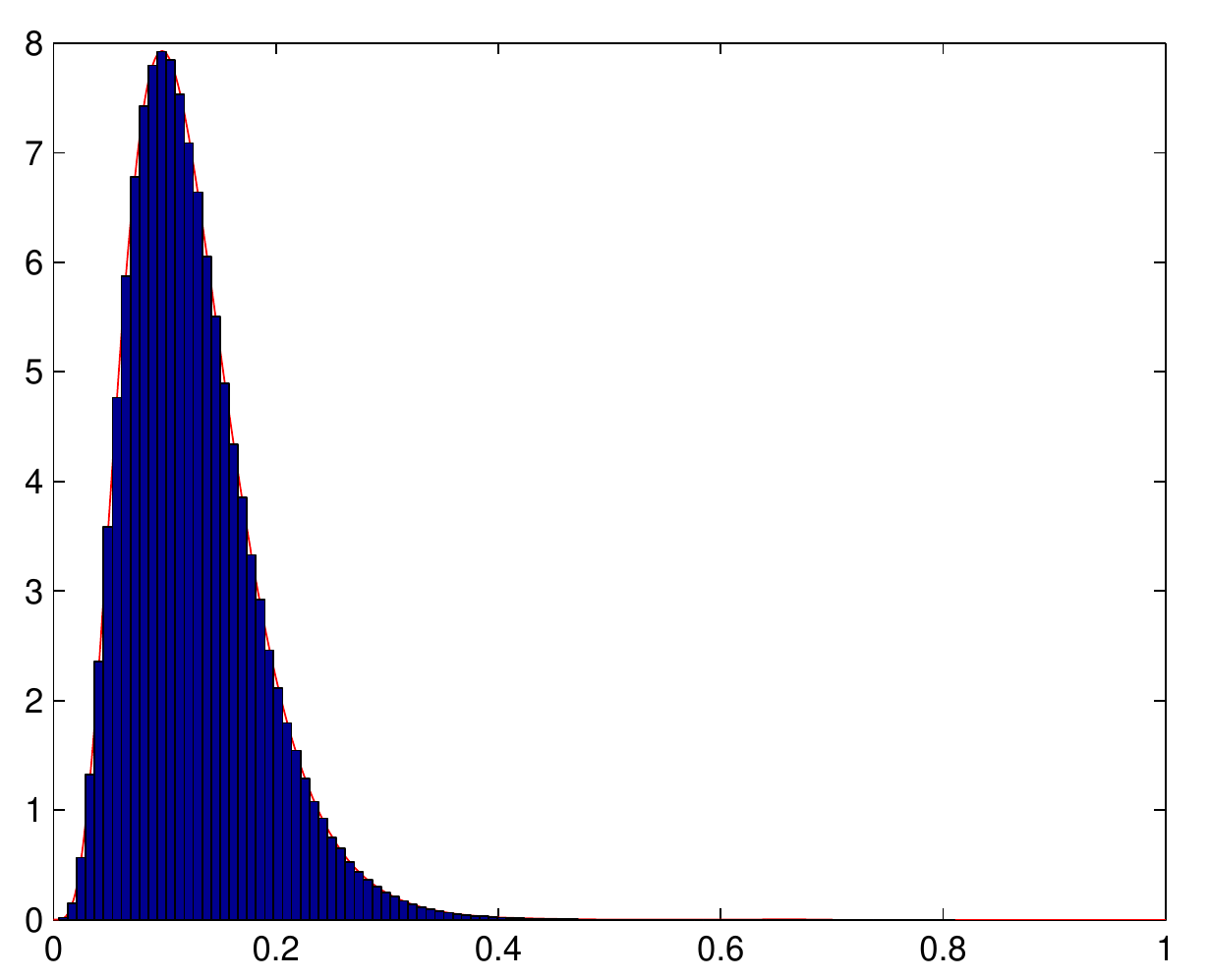}~\caption{\label{fig:histinvprec90} The histogram and the theoretical p.d.f. of $r_{jj}^2$ (red line) for $j=90$ and $10^6$ square channels of size $n=100$ using inverse precoder.}
	\end{center}
\end{figure}

In order to investigate the advantage of decoding at $\mathrm{B}$ over $\mathrm{E}$ under \aln{SIC} decoding algorithm, we define the following advantage ratio:
\begin{equation}~\label{def:advratioSIC}
\mbox{adv}_{\tiny \aln{SIC}}\triangleq r_{nn}^2({\bf I})/r_{nn}^2({\bf Q}),
\end{equation}
is called the advantage of $\mathrm{B}$ over $\mathrm{E}$ under \aln{SIC} attack. Since $r_{nn}^2({\bf I})=1$, the $\mbox{adv}_{\tiny \aln{SIC}} = 1/r_{nn}^2({\bf Q})$.
We now derive the exact distribution of the diagonal entries of the matrix ${\bf R}$ and specially the distribution and expected value of $r_{nn}^2({\bf Q})$. We cite the following theorem from~\cite{GuptaNagar99}:
\begin{theorem}~\label{th:chngvar}
Let ${\bf Q}$ be an $n\times n$ random full-rank matrix with probability density function $P$. If ${\bf Q} = {\bf O}{\bf R}$, where ${\bf R}$, $r_{jj}>0$ an upper triangular matrix and ${\bf O}$ is an orthogonal matrix, ${\bf O}{\bf O}^t ={\bf I}_n$, then ${\bf R}$ and ${\bf O}$ are independent and the probability density function of ${\bf R}$ is
\begin{equation}~\label{eq:changvar}
c_0\prod_{j=1}^nr_{jj}^{n-j}P\left({\bf R}{\bf R}^t\right),
\end{equation}
for a constant $c_0$.
\end{theorem}
Using the above theorem along with Theorem~\ref{th:qutionet}, we observe that the probability density function of ${\bf R}$ is proportional to
\begin{equation}~\label{eq:changvarfinal}
\prod_{j=1}^nr_{jj}^{n-j}\det\left({\bf I}_n+{\bf R}{\bf R}^t\right)^{-n}.
\end{equation}
We will make use of the following lemma from page $28$ of~\cite{Prasolov} in the proof of the next theorem.
\begin{lemma}~\label{lem:det}
Let ${\bf u}$ and ${\bf v}$ be columns of length $n$ and ${\bf A}$ be a square matrix of order $n$, then
\begin{itemize}
\item For a scalar $a$, we have
$$\det\left(\left[\begin{array}{cc}
{\bf A}&{\bf u}\\
{\bf v}^t&a
\end{array}
\right]\right)=a\det\left({\bf A}-{\bf u}{\bf v}^t/a\right).$$
\item The following equality holds:
$$\det\left({\bf A} + {\bf u} {\bf v}^t\right) = \det({\bf A}) + {\bf v}^t\mbox{adj}({\bf A}){\bf u},$$
where $\mbox{adj}({\bf A})$ denotes the adjoint of ${\bf A}$.
\end{itemize}
\end{lemma}
A random variable $v$ is said to have a beta distribution of the second type (beta prime distribution) $\mathcal{B}^{II}(a,b)$ if it has the following probability density function
$$v^{a-1}(1+v)^{-(a+b)}/\beta(a,b),~~~v>0,$$
where both $a$ and $b$ are non-negative and $\beta(a,b)$ is the beta function~\cite{GuptaNagar99}. The following theorem is now outstanding.
\begin{theorem}~\label{th:pdfRjj}
Let the matrices ${\bf Q}$, ${\bf O}$, and ${\bf R}$ be as in Theorem~\ref{th:chngvar}. Then $r_{jj}^2$ are independently distributed as $\mathcal{B}^{II}\left((n-j+1)/2,j/2\right)$, for $1\leq j\leq n$.
\end{theorem}
\begin{IEEEproof}
See Appendix~\ref{appen}.
\end{IEEEproof}
In Figs.~\ref{fig:histinvprec10}-\ref{fig:histinvprec90}, we show both the histogram of $r_{jj}^2$ and the probability density functions of $\mathcal{B}^{II}\left((n-j+1)/2,j/2\right)$ for different $j$'s equal to $10$, $40$, and $90$ for $10^6$ square channel matrices of size $n=100$. It is easy to check that these figures match perfectly suggesting the validity of Theorem~\ref{th:pdfRjj}.
\begin{figure}[htb]%
	\begin{center}%
		\includegraphics[width=7cm]{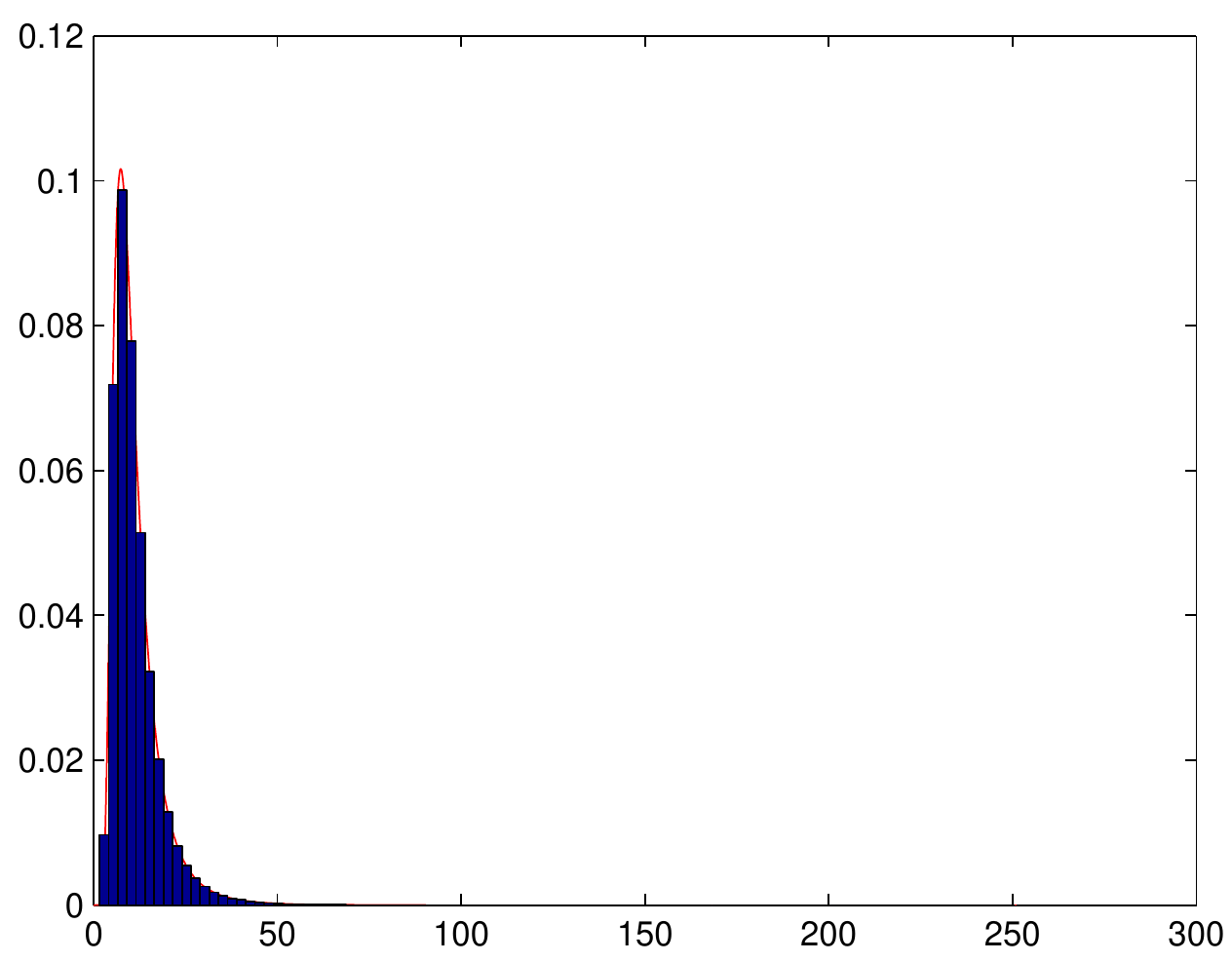}~\caption{\label{fig:histinvprec10} The histogram and the theoretical p.d.f. of $r_{jj}^2$ (red line) for $j=10$ and $10^6$ square channels of size $n=100$ using inverse precoder.}
	\end{center}
\end{figure}
The above calculation of a closed-form formula for the distribution of the diagonal elements of ${\bf R}$ in the \aln{QR} decomposition of ${\bf Q}$ enables us to find the $\mbox{adv}_{\tiny \aln{SIC}}$ of user $\mathrm{B}$ over $\mathrm{E}$ under \aln{SIC} decoding algorithm. Since the equivalent channel between users $\mathrm{A}$ and $\mathrm{B}$ is ${\bf I}$, the legitimate user can successfully decrypt the encrypted data. Next, we analyze the asymptotic behavior of $\mbox{adv}_{\tiny \aln{SIC}}$.
\begin{theorem}~\label{th:squareMMPLCInvPrecadv}
Let ${\bf H}_{n\times n}$ be the channel between $\mathrm{A}$ and $\mathrm{B}$ and ${\bf G}_{n \times n}$ be the channel between $\mathrm{A}$ and $\mathrm{E}$, both with i.i.d. elements each with distribution $\mathcal{N}_1$. Then, using the setting of \eqref{eq:InvPreCon} in \aln{MMPLC}, we get $\mbox{adv}_{\tiny \aln{SIC}}=\mathcal{O}\left(n\right)$.
\end{theorem}
\begin{IEEEproof}
See Appendix~\ref{appen}
\end{IEEEproof}
Note that the above result implies that $n/\omega(1)\leq\mbox{adv}_{\tiny \aln{SIC}}$.
On the other hand, since $r_{nn}^2({\bf Q})\geq \sigma^2_n({\bf Q})$, we get $\mbox{adv}_{\tiny \aln{SIC}}\leq \mbox{adv}_{\tiny \aln{ZF}}$,
which was itself upper bounded by $n^2$. This means that by using the computationally more complex (and of course non-linear) \aln{SIC} decoding algorithm, user $\mathrm{E}$ can gain more advantage over user $\mathrm{B}$ compared to when it uses \aln{ZF} linear receiver. However, the lower bound derived in Theorem~\ref{th:squareMMPLCInvPrecadv} implies that there is still security left even if user $\mathrm{E}$ employs a much stronger decoder than linear receivers. Next we show that user $\mathrm{E}$ basically cannot employ a maximum likelihood (\aln{ML}) decoder due to exponential dependency of the computational complexity of such algorithms to the dimension, $n$, of the Massive \aln{MIMO} channel.
%%%%%%%%%%%%%%%%%%%%%%%%%%%%%%%%%%%%%%%%%%%%%%%%%%%%%%%%%%%%%%%%%%%%%%%%%%%%%%%%%%%%%%%%%%%%%%%%%%%%%%%%%%%%%%%%
\subsection{Adversary with Sphere Decoder}~\label{AdversarywithSphereDecoder}
%%%%%%%%%%%%%%%%%%%%%%%%%%%%%%%%%%%%%%%%%%%%%%%%%%%%%%%%%%%%%%%%%%%%%%%%%%%%%%%%%%%%%%%%%%%%%%%%%%%%%%%%%%%%%%%%
Let us assume that user $\mathrm{E}$ has access to an optimal maximum likelihood (\aln{ML}) decoder such as a sphere decoder~\cite{ViB}. Since we found the closed forms of distributions of the upper triangular matrix ${\bf R}$ in the \aln{QR} decomposition of ${\bf Q}$, we can lower bound the complexity of a maximum likelihood (\aln{ML}) decoder such as the ones presented at~\cite{FP85,ViB} to find the encrypted data at user $\mathrm{E}$. In such algorithms the coordinates of the closest lattice vector are enumerated. Consider a basis $\{{\bf q}_1,\ldots, {\bf q}_n\}$ of the lattice $\Lambda_{\bf Q}$ with generator matrix ${\bf Q}$. Based on a heuristic analysis~\cite{NLLL}, one can see that the cost of enumeration
is lower bounded by $h_{n/2}$, which is approximately lower bounded by
\begin{equation}~\label{eq:lbsd}
h_{n/2}\geq2^{\Theta(n)}n^{\frac{n}{8}+o(n)}.
\end{equation}
Hence using a \aln{ML} decoder at user $\mathrm{E}$ in \aln{MMPLC} scheme is prohibited if the number of antennas $n$ is in the order of few hundreds, which is the case in Massive \aln{MIMO} setup.
\section{Summary and Directions for Future Work}~\label{sec7}
%%%%%%%%%%%%%%%%%%%%%%%%%%%%%%%%%%%%%%%%%%%%%%%%%%%%%%%%%%%%%%%%%%%%%%%%%%%%%%%%%%%%%%%%%%%%%%%%%%%%%%%%%%%%%%%%
%%%%%%%%%%%%%%%%%%%%%%%%%%%%%%%%%%%%%%%%%%%%%%%%%%%%%%%%%%%%%%%%%%%%%%%%%%%%%%%%%%%%%%%%%%%%%%%%%%%%%%%%%%%%%%%%
A Zero-Forcing (\aln{ZF}) attack has been presented for the massive multiple-input multiple-output \aln{MIMO} physical layer cryptosystem (\aln{MMPLC-13}) in~\cite{DG13v1}. A decoding advantage ratio has been defined and studied for \aln{ZF} linear receiver. It has been shown that this advantage tends to $1$ employing a singular value decomposition (\aln{SVD}) precoding approach at the legitimate transmitter and a \aln{ZF} linear receiver at the adversary. Our generalized upper bound on legitimate user to adversary \aln{ZF} decoding advantage suggests the complexity-based approach does not remove the needed linear limitation on the number of adversary antennas versus the number of the legitimate party antennas, that is also suffered by previous information-theoretic methods. 

The basic system of the updated \aln{MMPLC-17} in~\cite{DG13v2} is essentially the same as the one in~\cite{DG13v1} with different parameters for noise magnitude and constellation size as hardness conditions. In this paper, it has also been shown that the \aln{MMPLC-17} is cryptographically incorrect meaning it cannot deliver a unique message to the legitimate user. We also note that our \aln{ZF} analysis can be applied to the this general model, for any choice of parameters. In particular, we have shown that the user $\mathrm{B}$ has basically no \aln{ZF} \aln{MIMO} decoding advantage over user $\mathrm{E}$ when user $\mathrm{E}$ has the same (or bigger) number of receiving antennas.

We then turn our attention to the case, where all parties has the same number of antennas $n$. It is been proven under this circumstance, an advantage ratio in the order of $n^2$ is achievable. Although the proposed scheme would not be power efficient at all, one line of research is to design more power efficient precoders achieving maximum possible advantage ratio. If eavesdropper employs a stronger decoder algorithm such as a successive interference cancellation (\aln{SIC}), then the advantage ratio will be reduced to a constant fraction of $n$. Our positive result for the inverse precoder suggests that if the adversary is limited to have the same number of antennas as the legitimate parties, the complexity-based approach may provide practical security. This suggests the following questions: Can a security reduction from a worst-case standard lattice problem be given for this case? How does the practicality of the resulting scheme compare to existing physical-layer security schemes based on information-theoretic security arguments? Can the efficiency of those schemes be improved by the complexity-based approach?

\appendix~\label{appen}
{\bf Proof of Lemma~\ref{lem:noiseatEVE}:}
Note that $({\bf U}')^t{\bf e}'$ has the same distribution as ${\bf e}'$ since $({\bf U}')^t$ is orthogonal. Hence, $z_j$, the $j$-th coordinate of the vector ${\bf z} = ({\bf \Sigma}')^{-1}({\bf U'})^t{\bf e}'$ is distributed as $\mathcal{N}_{m^2\alpha^2/\sigma^2_j({\bf G})}$, for all $1\leq j\leq n_t$. We also note that $z_j$'s are independent with different variances.  We find the distribution of
\begin{equation}~\label{eq:innerproduct}
t_i = \langle{\bf v}_i',{\bf z}\rangle= \sum_{j=1}^{n_t} v'_{i,j}z_j,
\end{equation}
where $\langle{\bf v},{\bf w}\rangle\triangleq{\bf v}^t\cdot{\bf w}$ for a row vector ${\bf v}$ and a column vector ${\bf w}$. Since the linear combination of independent Gaussian random variables is again a Gaussian distributed random variable, $t_i$ in \eqref{eq:innerproduct} is distributed as
\begin{eqnarray}
\sum_{j=1}^{n_t} v'_{i,j}\mathcal{N}_{m^2\alpha^2/\sigma^2_j({\bf G})}\!\!\!\! &=&\!\!\!\! \mathcal{N}_{\sum_{j=1}^{n_t} |v'_{i,j}|^2m^2\alpha^2/\sigma^2_j({\bf G})}\label{eq:inner1}\\
\!\!\!\!&=&\!\!\!\! \mathcal{N}_{m^2\alpha^2\sum_{j=1}^{n_t} |v'_{i,j}|^2/\sigma^2_j({\bf G})}.\label{eq:inner2}
\end{eqnarray}
Since $\sigma^2_j({\bf G})\geq \sigma^2_{n_t}({\bf G})$, for all $1\leq j\leq n_t$, the random variable $t_i$ is distributed as $\mathcal{N}_{\sigma^2_{t_i}}$ with
\begin{equation}\label{eq:inn3}
\sigma^2_{t_i}\!=\!m^2\alpha^2\sum_{j=1}^{n_t} \!\frac{|v'_{i,j}|^2}{\sigma^2_j({\bf G})}\!\leq\!\frac{m^2\alpha^2}{\sigma^2_{n_t}({\bf G})}\sum_{j=1}^{n_t} |v'_{i,j}|^2=\!\frac{m^2\alpha^2}{\sigma^2_{n_t}({\bf G})},
\end{equation}
where the last equality holds because ${\bf V}'$ is orthogonal.

{\bf Proof of Theorem~\ref{th:PeZF}:}
Let $\mathcal{G}$ be the set of all channel matrices ${\bf G}$ such that
$$\sigma_{n_t}^2({\bf G})\geq n_r'\left(\left(1-1/\sqrt{y'}\right)^2-\varepsilon'\right).$$
Note that ${\bf G} \not\in \mathcal{G}$ with vanishing probability $o(1)$ as $n_t \rightarrow \infty$, by Theorem~\ref{th:sv}. We have:
\begin{eqnarray}
\mathbb{P}_{\tiny \aln{ZF}}[\mathrm{E}]\!\!\!\!&=&\!\!\!\!\!\mathbb{P}_{\tiny \aln{ZF}}[\mathrm{E}|{\bf G}\in\mathcal{G}]\mathbb{P}\left[{\bf G}\in\mathcal{G}\right]\!+\!\mathbb{P}_{\tiny \aln{ZF}}[\mathrm{E}|{\bf G}\notin\mathcal{G}]\mathbb{P}\left[{\bf G}\notin\mathcal{G}\right]\nonumber\\
\!\!\!&\leq&\!\!\!\! \mathbb{P}_{\tiny \aln{ZF}}[\mathrm{E}|{\bf G}\in\mathcal{G}]+\mathbb{P}\left[{\bf G}\notin\mathcal{G}\right]\nonumber\\
\!\!\!&\leq&\!\!\!\! n_t\mathbb{P}_{w \samp \mathcal{N}_{1}}\left[|w| < |\sigma_{n_t}({\bf G})|/(2 m \alpha) \right]+o(1)\nonumber\\
\!\!\!&\leq&\!\!\!\! n_t\exp\left(-\sigma^2_{n_t}({\bf G})/\left(8m^2\alpha^2\right)\right)+o(1)\nonumber\\
\!\!\!&\leq&\!\!\!\! n_t\exp\left(\frac{-n_r'((1-\sqrt{1/y'})^2-\varepsilon')}{8m^2\alpha^2}\right)+o(1),\label{eq:RHS0}
\end{eqnarray}
where in the first inequality we used $\mathbb{P}\left[{\bf G}\in\mathcal{G}\right]\!\leq\!1$ and
$$\mathbb{P}_{\tiny \aln{ZF}}[\mathrm{E}|{\bf G}\notin\mathcal{G}]\mathbb{P}\left[{\bf G}\notin\mathcal{G}\right]\leq\mathbb{P}\left[{\bf G}\notin\mathcal{G}\right],$$
the second inequality is true based on \eqref{Pe:AE} and Theorem~\ref{th:sv}, the third inequality uses the well-known upper bound $\exp\left(-x^2/2\right)$ for the tail of a Gaussian distribution and the last inequality follows from the definition of $\mathcal{G}$.  By letting~\eqref{eq:RHS0} be less than $\varepsilon$, the sufficient condition \eqref{eq:PeZF} can be obtained.

{\bf Proof of Proposition~\ref{prop:rectangularMMPLCGeneralPrecoderdisadv}:}
It is easy to see (please refer to 17-8, 7.(c) of~\cite{HBLA}) the two inequalities below hold for every ${\bf H}$, ${\bf G}$, and ${\bf P}$:
$$\left\{\begin{array}{l}
\sigma_{n_t}({\bf H}{\bf P}) \leq \sigma_1({\bf H})\sigma_{n_t}({\bf P}),\label{eq:ublsvproduct}\\
\sigma_{n_t}({\bf G}{\bf P}) \geq \sigma_{n_t}({\bf G})\sigma_{n_t}({\bf P}).\label{eq:lblsvproduct}
\end{array}
\right.$$
Hence, the advantage ratio \eqref{def:advratiogen} can be upper bounded as
\begin{equation}~\label{eq:advratioGeneralPrecoderMM-PLC}
\mbox{adv}_{\tiny \aln{ZF}} \leq \frac{\sigma^2_1({\bf H})\sigma^2_{n_t}({\bf P})}{\sigma^2_{n_t}({\bf G})\sigma^2_{n_t}({\bf P})}=\frac{\sigma^2_1({\bf H})}{\sigma^2_{n_t}({\bf G})}=\mbox{advup}_{\tiny \aln{ZF}}.
\end{equation}
Using Theorem~\ref{th:sv} for the numerator and the denominator of the RHS  of \eqref{eq:advratioGeneralPrecoderMM-PLC}, respectively, and $n_r/n_r' \rightarrow y/y'$, we get
$$\mbox{advup}_{\tiny \aln{ZF}} \rightarrow \frac{y(1+\sqrt{1/y})^2}{y'(1-\sqrt{1/y'})^2} = \left(\frac{\sqrt{y}+1}{\sqrt{y'}-1}\right)^2.$$
In the case $n_r'=n_r$ and $y=y'\rightarrow \infty$, the latter inequality gives $\mbox{advup}_{\tiny \aln{ZF}} \rightarrow 1$. Also, the inequality $\left(\sqrt{y}+1\right)^2/\left(\sqrt{y'}-1\right)^2 \geq 1$ implies (using $y=y'/\rho'$) that $\rho' \leq 1/(1-2/\sqrt{y'})^2$, and the RHS of the latter is $\leq 9$ for all $y' \geq 9$, which implies $\min(y',\rho') \leq 9$.

{\bf Proof of Theorem~\ref{th:qutionet}:} We prove each part separately
\begin{itemize} 
\item{} The proof follows the same lines of the proof of Theorem 4.2.1 of~\cite{GuptaNagar99}. The joint density of ${\bf G}$ and ${\bf H}$ is proportional to $$\mbox{etr}\left(-\left({\bf G}{\bf G}^t+{\bf H}{\bf H}^t\right)/2\right),$$ where $\mbox{etr}(\bf {M}) = \exp\left(\mbox{Tr}\left({\bf M}\right)\right)$ for a matrix ${\bf M}$. Changing the variable from ${\bf G}$ to ${\bf Q}$, it follows that the Jacobian $J\left({\bf G}\rightarrow{\bf Q}\right)$ (for a definition, please see page 12 of~\cite{GuptaNagar99}) is equal to $\left|\det({\bf H})\right|^n$ and hence the joint density of ${\bf H}$ and ${\bf Q}={\bf G}{\bf H}^{-1}$ is proportional to $$\mbox{etr}\left(-\left(\left({\bf I}+{\bf Q}{\bf Q}^t\right){\bf H}^t{\bf H}\right)/2\right)\left|\det({\bf H})\right|^n,$$ where to get the above equation we have used the fact that $\mbox{Tr}({\bf N}{\bf M}) = \mbox{Tr}({\bf M}{\bf N})$ for matrices ${\bf M}$ and ${\bf N}$. Now integrating out ${\bf H}$ (using multivariate Gamma integral (see equation (1.4.6) of~\cite{GuptaNagar99})) yields the density of ${\bf Q}$ proportional to \eqref{eq:denQ}.
\item{} We now study the eigenvalue distribution of the product matrix ${\bf W}={\bf Q}{\bf Q}^t$, which proves useful later in finding an achievable upper bound on the advantage ratio. By changing the variable from ${\bf Q}$ to ${\bf W}$, which introduces a factor of $\det({\bf W})^{1/2}$ (since $J\left({\bf Q}\rightarrow{\bf W}\right)=\det({\bf W})^{1/2}$) and then from ${\bf W}$ to its eigenvalues $\{\lambda_j\}$ and the eigenvectors, for $1\leq j\leq n$, we see that the joint eigenvalue distribution has the explicit functional form proportional to
\begin{equation}~\label{eq:pdflambda1}
\det({\bf W})^{-\frac{1}{2}}\prod_{\ell=1}^n\frac{1}{\left(1+\lambda_\ell\right)^{n}}\prod_{1\leq j < k \leq n}(\lambda_k-\lambda_j),
\end{equation}
for $\lambda_\ell\geq0$.
Since $\det({\bf W})^{-\frac{1}{2}}=\prod_{\ell=1}^n\lambda_\ell^{-\frac{1}{2}}$, the above equation \eqref{eq:pdflambda1} and hence the eigenvalue distribution of ${\bf W}$ is proportional to
\begin{equation}~\label{eq:pdflambda}
\prod_{\ell=1}^n\lambda_\ell^{-\frac{1}{2}}/\left(1+\lambda_\ell\right)^{n}\prod_{1\leq j < k \leq n}(\lambda_k-\lambda_j).
\end{equation}
By further changing the variable $w_\ell=\frac{1}{1+\lambda_\ell}$, for $1\leq \ell\leq n$, the joint probability density function \eqref{eq:pdflambda} is proportional to
\begin{equation}~\label{eq:cv}
\prod_{\ell=1}^nw_{\ell}^{-\frac{1}{2}}\left(1-w_\ell\right)^{-\frac{1}{2}}\prod_{1\leq j < k \leq n}(w_k-w_j),
\end{equation}
where in \eqref{eq:cv} the exponent $-1/2$ of $w_{\ell}$ was in fact deduced from $n-2+1/2-(n-1)$, for which
the term $n$ is from $1/(1+\lambda_\ell)^n$, the term $-2$ contributed from the Jacobian of the transformation from $\lambda_{\ell}$ to $w_{\ell}$, for $1\leq \ell\leq n$, the term $1/2$ has appeared due to $\lambda_\ell^{-1/2}$, and finally $-(n-1)$ is from the multiplications of the denominators of the second term in \eqref{eq:pdflambda} as
\begin{eqnarray}
\prod_{1\leq j < k \leq n}(\lambda_k-\lambda_j)&=&\prod_{1\leq j < k \leq n}\left(1/w_k-1/w_j\right)\nonumber\\
&=&\prod_{\ell=1}^nw_{\ell}^{-(n-1)}\prod_{1\leq j < k \leq n}(w_k-w_j).\nonumber
\end{eqnarray}
With this substitution, it is now easy to check that $0\leq w_\ell \leq 1$, since $\lambda_\ell\geq0$ (due to non-singularity of ${\bf Q}$ and positive definiteness of ${\bf W}$) and also $w_1\leq \cdots \leq w_n$ because $\lambda_1\geq\cdots\geq \lambda_n$.
\end{itemize}
{\bf Proof of Theorem~\ref{th:inverseprecoderachievability}}
We compute the probability that the $\mbox{adv}_{\tiny \aln{ZF}}$ be less than a polynomial function $G(n)$, $\mathbb{P}\left[\mbox{adv}_{\tiny \aln{ZF}}\leq G(n)\right]$. Based on the definition of the $\mbox{adv}_{\tiny \aln{ZF}}$, we get
\begin{eqnarray}
\mathbb{P}\left[\mbox{adv}_{\tiny \aln{ZF}} \leq G(n)\right]& = & \mathbb{P}\left[\sigma^2_n({\bf I}_n)/\sigma^2_n\left({\bf G}{\bf H}^{-1}\right)\leq G(n)\right]\nonumber\\
&=& \mathbb{P}\left[\sigma^2_n\left({\bf Q}\right)\geq1/G(n)\right]\nonumber\\
&=& \mathbb{P}\left[\lambda_n\left({\bf W}\right) \geq 1/G(n)\right] \label{eq:int1}
\end{eqnarray}
Let us now define
\begin{equation}~\label{def:w}
w\triangleq G(n)/(1+G(n)),
\end{equation}
hence, we get:
\begin{eqnarray}
\mathbb{P}\left[\lambda_n\geq 1/G(n)\right]\!\!\!\!& = & \!\!\!\!\mathbb{P}\left[1/w_n-1\geq1/G(n)\right]=\mathbb{P}[w_n\leq w] \nonumber\\
\!\!\!\!& = & \!\!\!\!\mathbb{P}[w_1\leq w,\ldots,w_n\leq w]\label{eq:integral0}\\
\!\!\!\!& = & \!\!\!\!\int_0^w\cdots\int_0^wc\prod_{\ell=1}^nw_{\ell}^{-\frac{1}{2}}\left(1-w_\ell\right)^{-\frac{1}{2}}\nonumber\\
\!\!\!\!&  &
\prod_{1\leq j < k \leq n}(w_k-w_j) dw_1 \cdots dw_n,\label{eq:integral1}\\
\!\!\!&\leq&\!\!\!\!cw^{\frac{n(n-1)}{2}}\int_0^1\cdots\int_0^1\prod_{\ell=1}^ny_{\ell}^{-\frac{1}{2}}\left(1-y_\ell\right)^{-\frac{1}{2}}\nonumber\\
\!\!\!\!&  & \prod_{1\leq j < k \leq n}(y_k-y_j) dy_1 \cdots dy_n,\label{eq:integral2}
\end{eqnarray}
for a constant $c$ (independent of $n$) where \eqref{eq:integral0} is true because of the ascending order in $w_\ell$ and \eqref{eq:integral2} is obtained based on the change of variable from $w_\ell$ to $y_\ell=w_\ell/w$ and the fact that
\begin{equation}~\label{eq:ineq}
\left(1-w_\ell\right)^{-\frac{1}{2}}=\left(1-wy_\ell\right)^{-\frac{1}{2}}\leq \left(1-y_\ell\right)^{-\frac{1}{2}}w^{-\frac{1}{2}},
\end{equation}
as $w\leq1$. In particular, $w<1$  based on its definition in \eqref{def:w} and $\lim_{G(n)\rightarrow\infty}w = 1$. Note that $w^{n(n-1)/2}$ in \eqref{eq:integral2} follows from the change of variable in $\prod_{1\leq j < k \leq n}(w_k-w_j)$ as there are exactly $n(n-1)/2$ elements in this multiplication and the Jacobian $w^n$ got canceled by two $w^{-n/2}$'s in $w_{\ell}^{-1/2}$ and the inequality in \eqref{eq:ineq}.
The last term in \eqref{eq:integral2} equals $s_n\left(-1/2,-1/2,1/2\right)$. Hence by substituting \eqref{eq:SelbergInt} into \eqref{eq:integral2} and then \eqref{eq:int1}, it follows that
\begin{eqnarray}
\mathbb{P}\left[\mbox{adv}_{\tiny \aln{ZF}} < G(n)\right]& \leq & cw^{\frac{n^2-n}{2}}s_n\left(-1/2,-1/2,1/2\right)\nonumber\\
&=& c'w^{\frac{n^2-n}{2}},\label{eq:RHS}
\end{eqnarray}
where $c'=cs_n\left(-1/2,-1/2,1/2\right)$.
We claim that $c'=1$. To see that, it is easy to plug in $w\rightarrow1$ in the integrations \eqref{eq:integral0}-\eqref{eq:integral2} and note that the inequality in \eqref{eq:ineq} becomes equality for $w\rightarrow1$. We get $\mathbb{P}[w_1\leq 1,\ldots,w_n\leq 1] = c'(1)^{(n(n-1)/2}$. On the other hand, since $w_\ell\leq 1$, for $1\leq \ell \leq n$, $\mathbb{P}[w_1\leq 1,\ldots,w_n\leq 1] = 1$, which implies that $c'=1$ or equivalently $c^{-1} = s_n\left(-1/2,-1/2,1/2\right)$.
Therefore, we have,
\begin{eqnarray}
\mathbb{P}\left[\mbox{adv}_{\tiny \aln{ZF}} < G(n)\right] &=& \left(1/(1+1/G(n))\right)^{(n^2-n)/2}\nonumber\\ &\leq& \exp(-(n^2-n)/(4\cdot G(n))),~\label{eq:limitfinal}
\end{eqnarray}
where in the last step we used the inequality $1+2x \geq \exp(x)$ for $0 \leq x \leq 1/2$, and that $G(n) \geq 1$ for sufficiently large $n$.
We now distinguish between three cases: ({\em i}) if $G(n) = c_Gn$ for a constant $c_G$. As $n\rightarrow\infty$, then \eqref{eq:limitfinal} goes to $0$. ({\em ii}) if $G(n) = c_Gn^2$ for a constant $c_G$. As $n\rightarrow\infty$, then \eqref{eq:limitfinal} goes to $e^{-1/c_G}$. And finally ({\em iii}) if $G(n) = c_Gn^3$ for a constant $c_G$, then by letting $n\rightarrow\infty$, we get $\left(G(n)/(1+G(n))\right)^{(n^2-n)/2}\rightarrow1$. %Fig.~\ref{fig:fig1} shows the above cases for $0\leq n \leq 100$.
The proof is now complete by taking the second case and verifying that the right hand side of~(\ref{eq:limitfinal}) is $\leq \varepsilon$ when $G(n) \leq \frac{1}{4 \log(1/\varepsilon)} \cdot (n^2-n)$. % the above and $G(n)=c_G'$ with $c_G'>c_G$ and $\varepsilon = 1-e^{-\frac{1}{c_G}}$.
\begin{figure}[htb]%
	\begin{center}%
		\includegraphics[width=7cm]{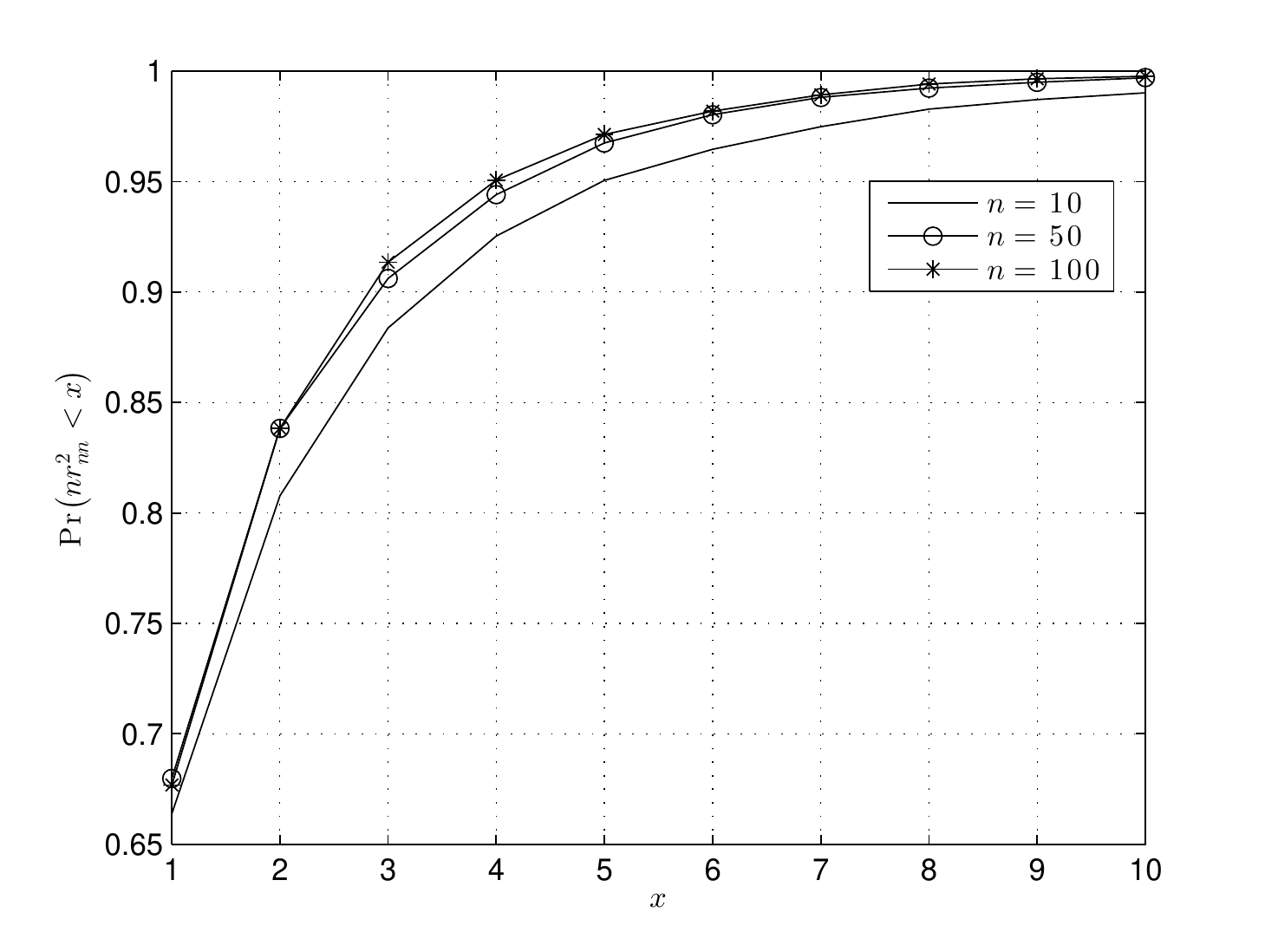}~\caption{\label{fig:Pr1} The numerical values of $\mathbb{P}\left[nr_{nn}^2({\bf Q})< x\right]$ for different dimensions $n=10$, $50$, and $100$ for $10000$ square channels of size $n=100$ using inverse precoder.}
	\end{center}
\end{figure}
	
{\bf Proof of Theorem~\ref{th:pdfRjj}:} Let us first find $\det\left({\bf I}_n + {\bf R}{\bf R}^t\right)$. Since ${\bf R}$ is an upper triangular matrix, it can be written as:
$${\bf R} = \left[
\begin{array}{cc}
{\bf R}_{11}& {\bf r}\\
{\bf 0}^t & r_{nn}
\end{array}\right].$$
It turns out that $\det\left({\bf I}_n + {\bf R}{\bf R}^t\right)$ can be expanded as what is given at the top of next page,
\begin{figure*}[!t]
	% ensure that we have normalsize text
	\normalsize
	% Store the current equation number.
	\setcounter{MYtempeqncnt}{\value{equation}}
	% Set the equation number to one less than the one
	% desired for the first equation here.
	% The value here will have to changed if equations
	% are added or removed prior to the place these
	% equations are referenced in the main text.
	\setcounter{equation}{38}
	\begin{eqnarray}
	\det\left({\bf I}_n + {\bf R}{\bf R}^t\right)
	&=&\det\left(\left[\begin{array}{cc}
	{\bf I}_{n-1}& {\bf 0}\\
	{\bf 0}^t & 1
	\end{array}\right]
	+\left[\begin{array}{cc}
	{\bf R}_{11}& {\bf r}\\
	{\bf 0}^t & r_{nn}
	\end{array}\right]
	\left[\begin{array}{cc}
	{\bf R}^t_{11}& {\bf 0}\\
	{\bf r}^t & r_{nn}
	\end{array}\right]\right)\nonumber\\
	&=&\det\left(\left[\begin{array}{cc}
	{\bf I}_{n-1}& {\bf 0}\\
	{\bf 0}^t & 1
	\end{array}\right]
	+\left[\begin{array}{cc}
	{\bf R}_{11}{\bf R}_{11}^t + {\bf r}{\bf r}^t& r_{nn}{\bf r}\\
	r_{nn}{\bf r}^t & r_{nn}^2
	\end{array}\right]\right)=
	\det\left(\left[\begin{array}{cc}
	{\bf I}_{n-1}+{\bf R}_{11}{\bf R}_{11}^t+{\bf r}{\bf r}^t& r_{nn}{\bf r}\\
	r_{nn}{\bf r}^t & 1+r^2_{nn}
	\end{array}\right]\right)\nonumber\\
	&=&
	\det\left(\left(1+r_{nn}^2\right)\left({\bf I}_{n-1}+{\bf R}_{11}{\bf R}_{11}^t+{\bf r}{\bf r}^t\right)-r_{nn}^2{\bf r}{\bf r}^t\right)\label{eq:rel1}\\
	&=&\left(1+r_{nn}^2\right)\det\left({\bf I}_{n-1}+{\bf R}_{11}{\bf R}_{11}^t+\left(1-r_{nn}^2(1+r_{nn}^2)^{-1}\right){\bf r}{\bf r}^t\right)\nonumber\\
	&=&\left(1\!+\!r_{nn}^2\right)\det\left({\bf I}_{n-1}\!+\!{\bf R}_{11}{\bf R}_{11}^t\right)\det\left({\bf I}_{n-1}+(1+r_{nn}^2)^{-1}\left({\bf I}_{n-1}+{\bf R}_{11}{\bf R}_{11}^t\right)^{-1}{\bf r}{\bf r}^t\right)\nonumber\\
	&=& \left(1+r_{nn}^2\right)\det\left({\bf I}_{n-1}+{\bf R}_{11}{\bf R}_{11}^t\right)\left(1+(1+r_{nn}^2)^{-1}{\bf r}^t\left({\bf I}_{n-1}+{\bf R}_{11}{\bf R}_{11}^t\right)^{-1}{\bf r}\right),\label{eq:rel2}
	\end{eqnarray}
	% Restore the current equation number.
	\setcounter{MYtempeqncnt}{\value{equation}}
	\setcounter{equation}{\value{MYtempeqncnt}}
	% IEEE uses as a separator
	\hrulefill
	% The spacer can be tweaked to stop underfull vboxes.
	\vspace*{4pt}
\end{figure*}
where \eqref{eq:rel1} and \eqref{eq:rel2} are obtained based on the first and the second parts of Lemma~\ref{lem:det}, respectively. In the latter case, we choose
${\bf A} = {\bf I}_{n-1}$,
$${\bf u} = (1+r_{nn}^2)^{-1}\left({\bf I}_{n-1}+{\bf R}_{11}{\bf R}_{11}^t\right)^{-1}{\bf r},$$
and ${\bf v} = {\bf r}$ and the facts that $\det\left({\bf I}_{n-1}\right)=1$ and $\mbox{adj}\left({\bf I}_{n-1}\right)={\bf I}_{n-1}$. Substituting \eqref{eq:rel2} into \eqref{eq:changvarfinal}, the joint density of ${\bf R}_{11}$, ${\bf r}$, and $r_{nn}^2$ is
$$J\left({\bf R}_{11}, {\bf r}, r_{nn}^2\right) = J_1\left({\bf R}_{11}\right)J_2\left(r_{nn}^2\right)J_3\left({\bf r}|{\bf R}_{11},r_{nn}^2\right),$$
where $J_1\left({\bf R}_{11}\right)$ is defined as
$$c_1\prod_{j=1}^{n-1}r_{jj}\prod_{j=1}^{n-1}r_{jj}^{(n-1)-j}\det\left({\bf I}_{n-1}+{\bf R}_{11}{\bf R}_{11}^t\right)^{-(n-1)},$$
$$J_2\left(r_{nn}^2\right)\triangleq c_2 \left(r_{nn}^2\right)^{1-1}\left(1+r^2_{nn}\right)^{-1-n/2)},$$
and $J_3\left({\bf r}|{\bf R}_{11},r_{nn}^2\right)$ is defined as
$$c_3\left(1+r^2_{nn}\right)^{-n/2+1}\det\left({\bf I}_{n-1}+{\bf R}_{11}{\bf R}_{11}^t\right)^{-1}$$
$$\left(1+(1+r_{nn}^2)^{-1}{\bf r}^t\left({\bf I}_{n-1}+{\bf R}_{11}{\bf R}_{11}^t\right)^{-1}{\bf r}\right)^{-n},$$
for appropriate constants $c_1$, $c_2$, and $c_3$. It is now easy to see that $J_2$ is proportional to a beta distribution of second type as $\mathcal{B}^{II}\left(1/1,n/2\right)$.
By further changing the variables:
$$\left\{
\begin{array}{l}
r'_n = r_{nn}^2,\\
{\bf t}' = (1+r_{nn}^2)^{\frac{-1}{2}}{\bf r}^t\left({\bf I}_{n-1}+{\bf R}_{11}{\bf R}_{11}^t\right)^{\frac{-1}{2}},
\end{array}
\right.$$
with the Jacobian
$$2\left(r'_n\right)^{\frac{-1}{2}}\left(1+r'_n\right)^{\frac{n-1}{2}}\det\left({\bf I}_{n-1}+{\bf R}_{11}{\bf R}_{11}^t\right)^{\frac{1}{2}},$$
we get that $r_{nn}^2$ is independent of ${\bf R}_{11}$, which itself has the same distribution as ${\bf R}$ with $n$ replaced by $n-1$. By recursively decomposing the joint distribution $J$ and its independent components $J_1$, $J_2$, and $J_3$, we further find the distributions of the other $r_{jj}^2$ for $n-1\leq j \leq 1$ as beta distributions of the second type as $\mathcal{B}^{II}\left((n-j+1)/2,j/2\right)$.

{\bf Proof of Theorem~\ref{th:squareMMPLCInvPrecadv}:}
We start by computing the $\mathbb{P}\left[r_{nn}^2\leq n/\omega(1)\right]$. We have the following Chebyshev's inequality:
\begin{equation}\label{eq:cheb1}
\mathbb{P}\left[\frac{r_{nn}^2-\mathbb{E}\left[r_{nn}^2\right]}{\sqrt{\mathbb{V}\left[r_{nn}^2\right]}}\leq t\right]\geq1-\frac{1}{t^2}.
\end{equation}
Since $r_{nn}^2$ is distributed based on $\mathcal{B}^{II}\left(\frac{1}{2},\frac{n}{2}\right)$, it follows that
\begin{equation}\label{eq:exprnn}
\mathbb{E}\left[r_{nn}^2\right]=\frac{1}{2}/\left(\frac{n}{2}-\frac{1}{2}\right) = 1/(n-1) =\mathcal{O}\left(1/n\right).
\end{equation}
and
\begin{equation}\label{eq:varrnn}
\mathbb{V}\left[r_{nn}^2\right]=\frac{(1/2)\left(1/2+n/2-1\right)}{\left(n/2-1\right)\left(n/2-2\right)^2} = \mathcal{O}\left(1/n^2\right).
\end{equation}
Substituting these into \eqref{eq:cheb1}, we get
\begin{equation}\label{eq:cheb2}
\mathbb{P}\left[(r_{nn}^2-c_e/n)/(c_v/n)\leq t\right]\geq1-1/t^2,
\end{equation}
for constants $c_e$ and $c_v$ independent of $n$. The above inequality is equivalent to $\mathbb{P}\left[r_{nn}^2\leq(tc_v+c_e)/n\right]\geq1-1/t^2$.
By letting $tc_v+c_e=\omega(1)$, we get $\mathbb{P}\left[r_{nn}^2\leq\omega(1)/n\right]\geq 1-o(1)$.
On the other hand $\mathbb{P}\left[r_{nn}^2\leq\omega(1)/n\right] = \mathbb{P}\left[1/r_{nn}^2\geq n/\omega(1)\right]$, which completes the proof. See Fig.~\ref{fig:Pr1} for a plot of $1\leq x \leq 10$ versus $\mathbb{P}\left[nr_{nn}^2< x\right]$.

\end{document}